\documentclass[aps,twocolumn,prd,showpacs,showkeys,preprintnumbers,superscriptaddress,nobibnotes,floatfix,longbibliography]{revtex4-1}

\pdfoutput=1
\usepackage{amsmath,amsfonts,amssymb,mathrsfs,graphicx,color,longtable}
\usepackage{hyperref}
\usepackage{graphicx}
\usepackage{amsfonts,amsmath,amssymb,bm}
\usepackage{array}
\usepackage{color}
\usepackage{enumitem}

\newcommand{\ie}{{\it i.e.}}

\newcommand{\eg}{{\it e.g.}}

\newcommand{\eq}{Eq.}

\newcommand{\equ}[1]{\eq~(\ref{equ:#1})}

\graphicspath{{figures/}}

\begin{document}


\title{Echo Technique to Distinguish Flavors of Astrophysical Neutrinos}

\author{Shirley Weishi Li}
\email{Now at SLAC National Accelerator Laboratory, \\ shirleyl@slac.stanford.edu}
\affiliation{Center for Cosmology and AstroParticle Physics (CCAPP), Ohio State University, Columbus, OH 43210}
\affiliation{Department of Physics, Ohio State University, Columbus, OH 43210}

\author{Mauricio Bustamante}
\email{Now at Niels Bohr Institute, mbustamante@nbi.ku.dk}
\affiliation{Center for Cosmology and AstroParticle Physics (CCAPP), Ohio State University, Columbus, OH 43210}
\affiliation{Department of Physics, Ohio State University, Columbus, OH 43210}

\author{John F. Beacom}
\email{beacom.7@osu.edu}
\affiliation{Center for Cosmology and AstroParticle Physics (CCAPP), Ohio State University, Columbus, OH 43210}
\affiliation{Department of Physics, Ohio State University, Columbus, OH 43210}
\affiliation{Department of Astronomy, Ohio State University, Columbus, OH 43210}

\date{April 18, 2019}

\begin{abstract}
The flavor composition of high-energy astrophysical neutrinos is a rich observable.  However, present analyses cannot effectively distinguish particle showers induced by $\nu_e$ versus $\nu_\tau$.  We show that this can be accomplished by measuring the intensities of the delayed, collective light emission from muon decays and neutron captures, which are, on average, greater for $\nu_\tau$ than for $\nu_e$.  This new technique would significantly improve tests of the nature of astrophysical sources and of neutrino properties.  We discuss the promising prospects for implementing it in IceCube and other detectors.
\end{abstract}

\maketitle


{\bf Introduction.---}
High-energy astrophysical neutrinos, long sought, were recently discovered by the IceCube Collaboration~\cite{Aartsen:2013bka, Aartsen:2013jdh, Aartsen:2013eka, Aartsen:2014gkd, Aartsen:2015ita, Aartsen:2015rwa}.  Their energy spectrum provides important clues about extreme astrophysical sources as well as neutrino properties at unexplored energies.  However, pressing mysteries remain.

Exploiting the flavor composition --- the ratios of the fluxes of $\nu_e + \bar{\nu}_e$, $\nu_\mu + \bar{\nu}_\mu$, and $\nu_\tau + \bar{\nu}_\tau$ to the total flux --- offers crucial additional clues.  In the nominal scenario, a composition of $\left( \frac{1}{3} : \frac{2}{3} : 0 \right)_\text{S}$ at the source is transformed by neutrino vacuum mixing to $\left( \frac{1}{3} : \frac{1}{3} : \frac{1}{3} \right)_\oplus$ at Earth~\cite{Learned:1994wg, Athar:2000yw}.  Even for arbitrary flavor composition at the source, the maximal range of flavor composition at Earth with only standard mixing is surprisingly narrow~\cite{Bustamante:2015waa}, making deviations sensitive indicators of new physics~\cite{Beacom:2002vi, Barenboim:2003jm, Beacom:2003nh, Beacom:2003zg, Beacom:2003eu, Xing:2006uk, Lipari:2007su, Pakvasa:2007dc, Esmaili:2009dz, Lai:2009ke, Bazo:2009en, Choubey:2009jq, Bustamante:2010nq, Bustamante:2010bf, Baerwald:2012kc, Laha:2013lka, Chen:2013dza, Mena:2014sja, Ng:2014pca, Xu:2014via, Fu:2014isa, Chen:2014gxa, Palomares-Ruiz:2015mka, Palladino:2015zua, Aartsen:2015ivb, Palladino:2015vna, Arguelles:2015dca, Palladino:2015uoa, Shoemaker:2015qul, Vincent:2016nut}.

So far, IceCube measurements of the flavor composition mostly separate muon tracks --- made primarily by charged-current (CC) $\nu_\mu+ \bar{\nu}_\mu$ interactions --- from particle showers --- made by all other interactions.  A significant limitation is their poor ability to distinguish between CC interactions of $\nu_e$ and $\nu_\tau$ (unless noted, $\nu_l$ refers to $\nu_l + \bar{\nu}_l$).

\begin{figure}[t]
    \begin{center}                  
        \includegraphics[width=\columnwidth,clip=true,trim = 0 0 0 0.55cm]{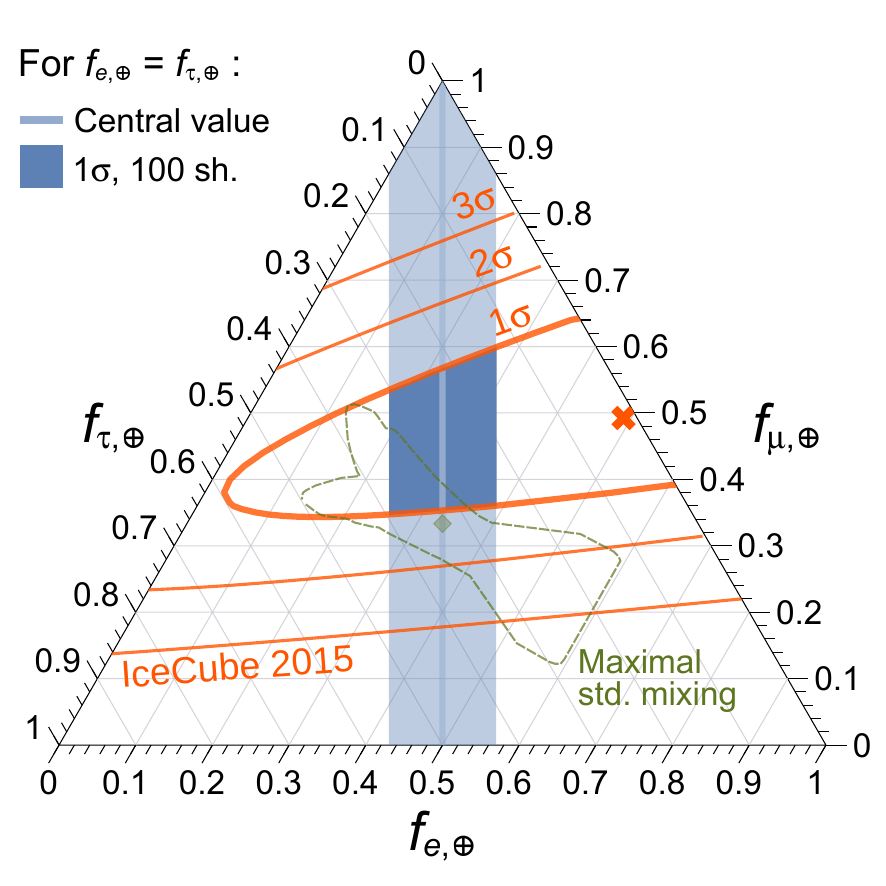}
        \caption{Flavor composition $f_{l,\oplus}$ ($l = e,\mu,\tau$) of astrophysical neutrinos at Earth.  Each axis is read parallel to its ticks.  Orange: the IceCube fit~\cite{Aartsen:2015ita}.  Blue: the expected precision of our proposed technique (for the case $f_{e,\oplus} = f_{\tau,\oplus}$), assuming 100 showers of 100~TeV with well-detected echoes.  Results for other energies are similar, but collecting 100 showers with echoes will likely require a larger detector than IceCube.  Green: the standard expectation~\cite{Learned:1994wg, Athar:2000yw} and maximal range with standard mixing~\cite{Bustamante:2015waa}.}
        \label{fig:flavor_triangle}
    \end{center}
    \vspace{-1em}
\end{figure}


{\bf Synopsis of the paper.---}
We propose a new technique to break this $\nu_e$-$\nu_\tau$ degeneracy, one that could work for a wider range of energies than existing ideas (Glashow resonance~\cite{Glashow:1960zz, Anchordoqui:2004eb, Bhattacharya:2011qu}, double pulses~\cite{Aartsen:2015dlt}, double bangs~\cite{Learned:1994wg}, and lollipops~\cite{DeYoung:2006fg}).

We introduce two new shower observables.  In showers, many low-energy muons and neutrons are produced; after delays, the muons decay and the neutrons capture.  We call the collective Cherenkov emission from the many independent decays and captures the {\it muon echo} and the {\it neutron echo}.  We show that the echoes are brighter for $\nu_\tau$-initiated than for $\nu_e$-initiated showers, which could allow them to be distinguished on a statistical basis.

Our focus is pointing out new observables to help solve the important problem of flavor identification.  The technical aspects of implementation require experimental expertise.  Nevertheless, in a preliminary evaluation, grounded in the measured properties of IceCube, we find the detection prospects promising.

Figure~\ref{fig:flavor_triangle} shows that the present $\nu_e$-$\nu_\tau$ degeneracy in IceCube elongates the contours of the measured flavor composition~\cite{Aartsen:2015ita}.  It also shows how detecting echoes could refine these measurements, probing the flavor composition better than the maximal range with standard mixing~\cite{Bustamante:2015waa}, which would lead to powerful conclusions.


{\bf High-energy neutrino signatures.---}
At present, IceCube identifies neutrino-initiated events only as tracks and showers, for which the Cherenkov light appears to emanate from approximate lines and spheres.  Tracks are caused by muons, which travel up to $\sim 10$~km in ice~\cite{Aartsen:2013jdh}, due to their low interaction and decay rates.  Showers are caused by all other neutrino-induced particles and extend only $\sim$ 10~m in ice~\cite{Aartsen:2013jdh}, due to the high interaction and decay rates of their constituent particles.

Neutrinos produce secondaries through deep-inelastic scattering~\cite{Gandhi:1995tf, Gandhi:1998ri, Connolly:2011vc}.  A neutrino interacts with a nucleon $N$ via the CC channel $\nu_l + N \rightarrow l + X$ or the neutral-current (NC) channel $\nu_l + N \rightarrow \nu_l + X$, where $l = e, \mu, \tau$, and $X$ represents hadrons.  A fraction $(1 - y)$ of the neutrino energy goes to the final-state lepton; the remaining fraction $y$ goes to the final-state hadrons.  The inelasticity distribution peaks at $y = 0$ and has an average $\langle y \rangle \approx 0.3$ at 100~TeV, for both $\nu$ and $\bar{\nu}$, CC and NC.

Tracks are produced by $\nu_\mu$ CC interactions plus 17\% of $\nu_\tau$ CC interactions where the tau decays to a muon~\cite{Agashe:2014kda}.  

Showers are produced by all other neutrino interactions.  For $\nu_e$ CC interactions, the electron- and hadron-initiated showers combine, and their sum energy equals the neutrino energy.  For $\nu_\tau$ CC interactions, the tau decays promptly, so again the showers combine (when the tau does not decay to a muon); the neutrino energy estimate is slightly biased because $\sim 25\%$ of its energy is lost to outgoing neutrinos from tau decay.  For NC interactions of all flavors, the hadron-initiated shower carries a fraction $y$ of the neutrino energy; because of the steeply falling neutrino spectrum, NC interactions are subdominant in the total shower spectrum~\cite{Beacom:2004jb}.  (This is also true for misidentified $\nu_\mu$ CC interactions that appear to be a shower event because the track is missed~\cite{Aartsen:2015ivb}.)

These points explain the basic features of the IceCube results in Fig.~\ref{fig:flavor_triangle}.  Because there are track events, the $\nu_\mu$ component of the flux must be nonzero; because there are shower events, the sum of the $\nu_e$ and $\nu_\tau$ components must be nonzero.  The similarity of $\nu_e$- and $\nu_\tau$-initiated events makes the contours nearly horizontal; the degeneracy is weakly broken because increasing the $\nu_\tau/\nu_e$ fraction increases the number of tracks and decreases the shower energies.  With present methods, improvement requires much larger exposure~\cite{Aartsen:2014njl, Bustamante:2015waa, Kowalski2018}.


{\bf Electromagnetic versus hadronic showers.---}
The key to our new method is understanding the low-energy physics underlying high-energy showers~\cite{Heitler1954, Matthews2005, Lipari:2008td, Lipari:2009zz, Rott:2012qb, Li:2014sea, Li:2015kpa, Li:2015lxa}.

When showers are developing, particles multiply in number while decreasing in energy.  An electromagnetic shower starts out with electrons, positrons, and gamma rays and stays composed predominantly of them; there is usually a small fraction of pions and nucleons produced by photonuclear processes.  A hadronic shower starts out with pions and nucleons, and then builds up a progressively larger fraction of electromagnetic particles as prompt $\pi^0 \rightarrow \gamma\gamma$ decays deplete $\sim 1/3$ of the remaining hadronic energy with each shower generation.

Shower development ends when the average particle energy is low enough that the particle- and energy-loss rates exceed the particle-production rates.  At that point, the most abundant particles in all showers are $\sim 100$-MeV electrons and positrons, which produce most of the prompt Cherenkov light.  Pions carry only $\sim 10$\% of the energy in hadronic showers and $\sim 1$\% in electromagnetic showers.  However, they are the key to separating electromagnetic and hadronic showers.


{\bf New shower observables.---}
At the end of shower development, charged pions come to rest by ionization; then $\pi^-$ capture on nuclei and $\pi^+$ decay to $\mu^+$.  The $\mu^+$ decay with a lifetime of 2.2~$\mu$s, producing $e^+$ with $\sim$ 35~MeV.  The collective Cherenkov light from these positrons is our first new observable: the {\it muon echo}.

Separately, neutrons lose energy by collisions until they reach thermal energy.  They eventually capture on hydrogen, with a timescale of $\sim $ 200~$\mu$s, producing 2.2~MeV gamma rays.  (In seawater, 33\% of neutrons capture on Cl; the emitted gamma rays have $8.6$~MeV~\cite{Tuli2016}, making the neutron echoes much more visible.)  The gamma rays Compton-scatter electrons to moderate energies, producing Cherenkov light.  This collective emission is our second new observable: the {\it neutron echo}.

\begin{figure}[t]
    \begin{center}                  
        \includegraphics[width=\columnwidth]{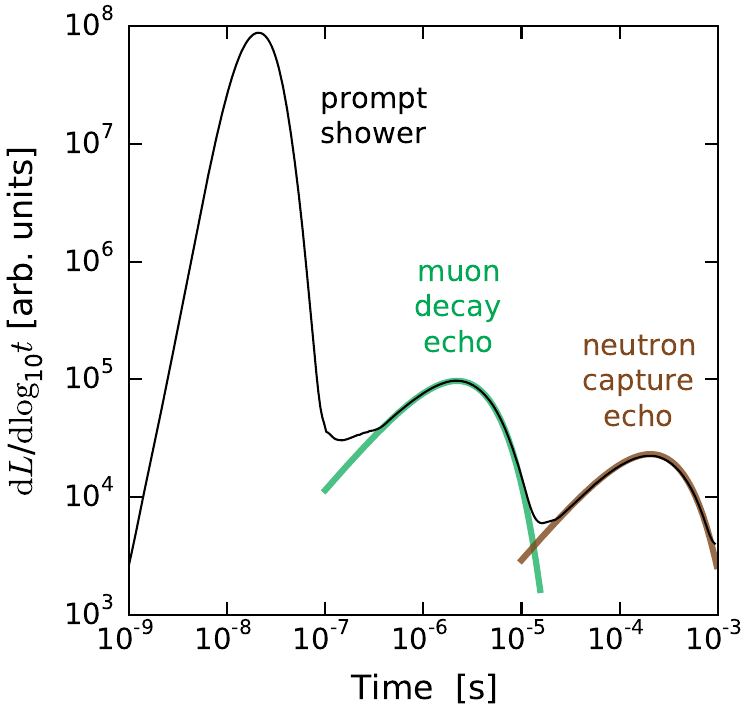}
        \caption{Time evolution of the light yield of a hadronic shower simulated with \texttt{FLUKA}, following injection of a 100-TeV charged pion.  The shaded bands are exponentials with the respective timescales.  For an electromagnetic shower of the same prompt energy, the echoes are $\sim$ 10 times smaller.}
        \label{fig:time_scale}
    \end{center}
    \vspace{-1em}
\end{figure}

We simulate showers and subsequent echoes using the \texttt{FLUKA} Monte Carlo software (version 2011.2c-4)~\cite{Ferrari2005, Battistoni2007}.  We inject high-energy electrons or positrons to simulate electromagnetic showers and charged pions to simulate hadronic showers.

Figure~\ref{fig:time_scale} shows the averaged time profile of a 100-TeV hadronic shower.  Because the features happen on very different timescales, it is appropriate to analyze their light yield $L$ in bins of log time.  Accordingly, we plot $\text{d}L/\text{d}\!\log t \propto t~\text{d}L/\text{d}t$; this makes the height of the curve proportional to its contribution to the integrated light yield.  The echo shapes are exponentials with the respective timescales.  The echoes are well separated from the prompt shower and from each other.

Figure~\ref{fig:time_scale} also shows that the echoes have low intensities: the muon echo has $\sim 3\times 10^{-3}$ of the prompt shower energy and the neutron echo has $\sim 6\times 10^{-4}$.  The first number results from the facts that 10\% of hadronic shower energy goes to pions, 10\% of those pions are $\pi^+$ that come to rest and decay, and 30\% of the pion decay energy goes to positrons from muon decays.  The second number results from the facts that there are about 10 times more neutron captures than muon decays, that the capture energy is about 20 times smaller, and that the Cherenkov efficiency is about 3 times smaller.

\begin{figure}[t]
    \begin{center}                  
        \includegraphics[width=\columnwidth]{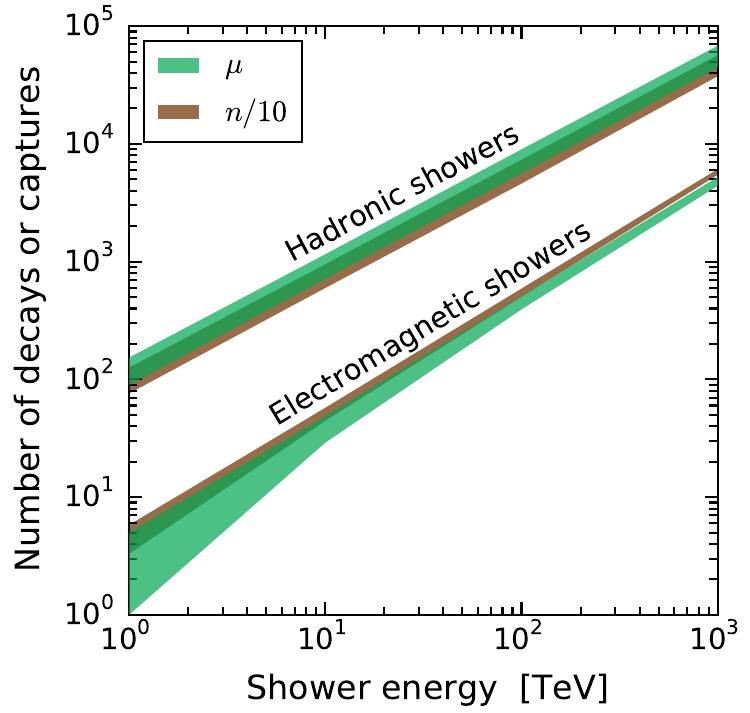}
        \caption{Numbers of muon decays and neutron captures (scaled down by 10) per shower, as a function of shower energy, for electromagnetic and hadronic showers simulated with \texttt{FLUKA}.  The bands show 1$\sigma$ intrinsic fluctuations.}
        \label{fig:N_mu}
    \end{center}
    \vspace{-1em}
\end{figure}


The points above carry over for electromagnetic showers, except for a crucial difference:  the pions carry only $\sim$ 1\% of the shower energy as opposed to $\sim$ 10\%.  Thus, the echo intensities are expected to be $\sim$ 10 times higher in hadronic showers than in electromagnetic showers.

Figure~\ref{fig:N_mu} shows that there are indeed about 10 times as many muon decays and neutron captures in hadronic showers.  This difference is much larger than the intrinsic fluctuations of these numbers.  Because the number of decays and captures, and, therefore, the light coming from them, grows linearly with shower energy, this factor-of-10 difference between electromagnetic and hadronic showers is present at all energies.  The yields may have an overall shift of up to a factor of 2 due to hadronic and nuclear uncertainties~\cite{Abe:2009aa, Bellini:2013pxa, Li:2014sea, Super-Kamiokande:2015xra}, but this can be calibrated by external measurements~\cite{Beacom:2003nk, Askins:2015bmb, Anghel:2015xxt, Fernandez:2016eux} or {\it in situ}.


{\bf Separating $\nu_e$ and $\nu_\tau$.---}
We now examine how echoes can be used to help identify the flavors of neutrino-induced showers.  In realistic neutrino interactions, the differences in the echoes are less stark than above.

Showers initiated by $\nu_e$ are mostly electromagnetic because the outgoing electron typically carries more energy than the final-state hadrons.  But showers initiated by $\nu_\tau$ are mostly hadronic because, in addition to the shower from the final-state hadrons, 67\% of tau decays are hadronic.  (NC showers are purely hadronic.)

We consider flavor separation at fixed shower energy, as opposed to fixed neutrino energy, to make contact with experiment.  We simulate neutrino interactions with appropriate energies to give $E_\text{sh} = 100$~TeV, including 10\% energy resolution~\cite{Aartsen:2013vja}.  For NC interactions, we mimic the final-state hadrons by directly injecting charged pions at the shower energy.

\begin{figure}[t]
    \begin{center}                  
        \includegraphics[width=\columnwidth]{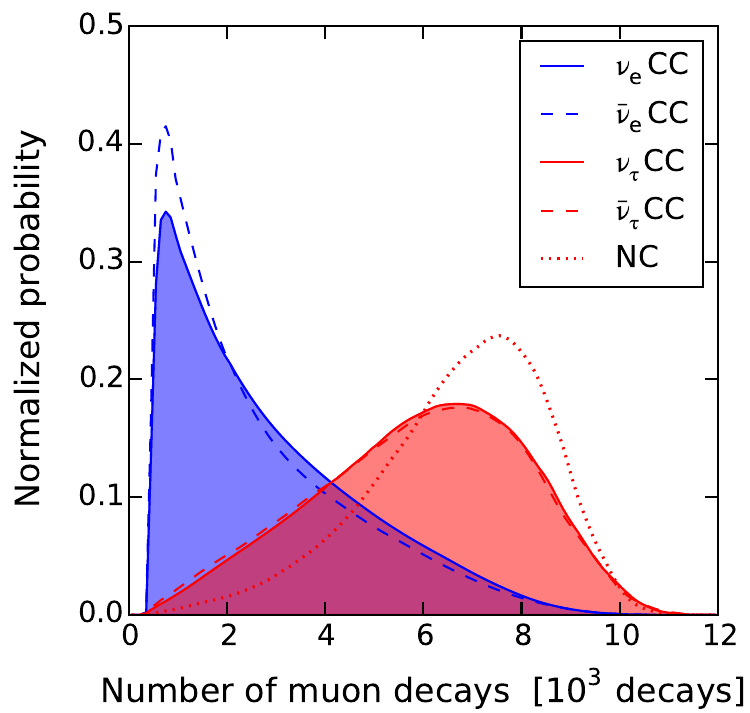}
        \caption{Probability distributions of the numbers of muon decays per shower (of energy 100~TeV) for different neutrino interaction channels, each normalized separately.}
        \label{fig:muon_decay_distribution}
    \end{center}
    \vspace{-1em}
\end{figure}

Figure~\ref{fig:muon_decay_distribution} shows how the numbers of muon decays per shower are distributed for different neutrino interaction channels.  As expected, $\nu_e$ CC showers produce fewer muons than $\nu_\tau$ CC showers.

The basics of the distributions in Fig.~\ref{fig:muon_decay_distribution} can be understood easily.  For pure electromagnetic showers, the peak would be at $\sim 500$ decays; it would be narrow because most pions are produced late in the shower and the fluctuations are mostly Poissonian.  For pure hadronic showers, the peak would be at $\sim 8000$ decays; it would be broad because there are large fluctuations in how much energy goes into $\pi^0$ in the first few shower generations.  The shapes shown in Fig.~\ref{fig:muon_decay_distribution} depend also on the $y$ distributions for neutrino interactions.  For $\nu_e$ CC events, the distribution is substantially broadened because the differential cross section $d\sigma/dy$, while peaked at $y=0$, has a substantial tail.  For $\nu_\tau$ CC events, there is a slight shift to the left, due to the 17\% of tau decays to muons.

The results in Fig.~\ref{fig:muon_decay_distribution} make it possible to distinguish $\nu_e$ and $\nu_\tau$ on a statistical basis.  We next estimate the sensitivity to flavor composition using the echoes from an ensemble of events, assuming well-detected echoes.

First, we use the results in Fig.~\ref{fig:muon_decay_distribution} to generate the muon decay distributions for each flavor, assuming an equal flux of $\nu_l$ and $\bar{\nu}_l$, and NC to CC event ratios consistent with a power-law spectral index of 2.5~\cite{Aartsen:2015ita}.  Next, for an assumed flavor composition, we randomly sample the number of muon decays for each shower in an ensemble of 100 showers of $E_\text{sh} = 100$~TeV.  Then, we treat the flavor composition $f_{e,\oplus}$ and $f_{\tau,\oplus}$ as free parameters ($f_{\mu,\oplus} = 1 - f_{e,\oplus} - f_{\tau,\oplus}$) and use an unbinned maximum-likelihood procedure to find their best-fit values.  We generate $10^3$ different realizations of the shower ensemble, and find the average best-fit values and uncertainties of $f_{e,\oplus}$ and $f_{\tau,\oplus}$.  Further details are in the Supplemental Material.

Figure~\ref{fig:flavor_triangle} shows the predicted sensitivity on $f_{e,\oplus}$ and $f_{\tau,\oplus}$, assuming equal $\nu_e$ and $\nu_\tau$ content, {\it i.e.}, a composition of the form $\left( x : 1-2x : x \right)_\oplus$, where $x$ varies in $\left[ 0, 0.5 \right]$.  The vertical shape of the band shows that the sensitivity to $f_{e,\oplus}$ and $f_{\tau,\oplus}$ does not depend on the $\nu_\mu$ content.  (Because our method is only weakly sensitive to $f_{\mu,\oplus}$, we suppress its uncertainty in the plot.)

Our results are conservative.  The sensitivity improves slightly with shower energy.  Assuming well-detected echoes, the sensitivity is comparable whether we use muon echoes only, neutron echoes only, or both.  It is also comparable, or better, for other choices of input parameters.  See the Supplemental Material for details.


{\bf Observability of the echoes.---}
Echo detection depends on how the echo light yield compares to that from ambient backgrounds and detector transients. These quantities are detector-dependent, and we use IceCube as a concrete example, either for itself or to guide the design of upgrades or future detectors.

The echoes are faint, but they are well localized, which enhances their visibility.  In space, like the parent shower, they are concentrated among only the few photomultiplier tubes (PMTs) on a single string that are closest to the neutrino interaction vertex~\cite{Kopper2013}.  In time, they occur $\sim 2.2$ $\mu$s and $\sim 200$ $\mu$s after the prompt shower.  These timescales require long-time data collection, made possible by the recent development of the HitSpooling technique, which can go to hours for infrequent events~\cite{Aartsen:2013nla, Aartsen:2016nxy}.  In direction, the shower light is beamed forward but the echo light is isotropic.  Light scattering makes the shower more isotropic and increases its duration~\cite{Aartsen:2013vja}, which could partially obscure the muon echo.

The total light yield of a shower in IceCube is $\sim 100$ detected photoelectrons (p.e.) per TeV~\cite{Aartsen:2013jdh}.  For 100~TeV, the muon echo in a hadronic shower is expected to yield $\sim 30$~p.e.\ (300 GeV) and the neutron echo $\sim 6$~p.e\ (60~GeV).  At 30 p.e., IceCube can easily trigger on neutrino events of that yield~\cite{Aartsen:2015xup}; the low efficiency at that energy shown in many analyses reflects the need to reject atmospheric muons.  For echoes, because the place and time of the preceding shower is known, IceCube could restrict attention to a small number of PMTs, allowing triggering on echoes with as few as 4 p.e.~\cite{Koepke:private, Aartsen:2016nxy}.  The minimum energy for our method is thus at least $\sim 10$~TeV.

Ambient backgrounds in IceCube do not eclipse the echoes.  For an average p.e.\ noise rate of $\sim 500$~Hz per PMT~\cite{Abbasi:2010vc}, the expected backgrounds in 2 and $200\ \mu$s are only $\sim 10^{-3}$ and $\sim 10^{-1}$ p.e.\ per PMT, respectively.  (Even with correlated noise, due to nuclear decays near the PMT, the backgrounds will be small in all but a few PMTs, and those will be identifiable~\cite{Larson2013}.)  And the cosmic-ray muon rate in IceCube is 3~kHz~\cite{Aartsen:2013jdh}, so the probability of a muon lighting up several specific PMTs in the short time between shower and echo is small.

A serious concern is the detector transient called afterpulsing, where a PMT registers late p.e.\ with total charge proportional to the initial signal (the shower) and with a time profile characteristic to the PMT.  For the IceCube PMTs, the muon echo will compete with an afterpulse feature of relative amplitude $\sim 10^{-2} E_\text{sh}$ near 2 $\mu$s~\cite{Abbasi:2010vc}; though larger than the echo, it is not overwhelmingly so.  Encouragingly, the neutron echo, though smaller, is late enough that afterpulsing seems to be negligible.
   
In summary, the prospects for observing echoes are promising, and they improve with shower energy.  Doing so may require changes in detector design or in PMT technology~\cite{Incandela:1987dh, Bristow02}; these considerations may shape the design of IceCube-Gen2~\cite{Aartsen:2014njl}, KM3NeT~\cite{Katz:2006wv, Adrian-Martinez:2016fdl}, and Baikal-GVD~\cite{Avrorin:2013sla}.  With multiple nearby PMTs~\cite{Resconi2013, Aartsen:2014njl, Adrian-Martinez:2016fdl}, it may be possible to reconstruct individual events, dramatically improving background rejection.  The final word on the observability of the echoes will come from detailed studies by the experimental collaborations.


{\bf Conclusions.---}
The rich phenomenology contained in the flavor composition of high-energy astrophysical neutrinos cannot be fully explored due to the difficulty of distinguishing showers initiated by $\nu_e$ versus $\nu_\tau$ in neutrino telescopes.  To break this degeneracy, we have introduced two new observables of showers: the delayed, collective light, or ``echoes,'' from muon decays and neutron captures.  This light reflects the size of the hadronic component of a shower, and it is stronger in $\nu_\tau$-initiated than $\nu_e$-initiated showers.

Figure~\ref{fig:flavor_triangle} shows the promise of our method.  With 100 showers with well-detected echoes, this would improve the separation of $\nu_e$ and $\nu_\tau$ by a factor of $\sim 9$ over present measurements~\cite{Aartsen:2015ita}.  That is comparable to the estimated sensitivity attainable with present techniques after more than 15 years of exposure of the next-generation detector IceCube-Gen2~\cite{Aartsen:2014njl,Kowalski2018}.  The observation of other flavor-specific event signatures --- Glashow resonance, double bangs, double pulses, and lollipops --- will further constrain the flavor composition.

The applications of tagging hadronic showers via muon and neutron echoes extend beyond flavor discrimination.  The technique could improve shower energy reconstruction, by folding in the probability of a shower being electromagnetic or hadronic.  And, at the considered energies, the echoes are shifted forward along the shower direction by $\sim 5$~m from the shower peak.  If this shift can be detected, it would improve the poor angular resolution~\cite{Aartsen:2013vja} of showers.

High-energy neutrino astronomy has just begun.  We are still learning the best ways to detect and analyze astrophysical neutrinos.  We should pursue all potentially detectable signatures, moving closer to finding the origins and properties of these ghostly messengers.

{\bf Note added:} Preliminary results from a search for echoes in IceCube data are very encouraging~\cite{Steuer:2017tca}. \\


\begin{acknowledgments}
We thank Carlos Arg\"uelles, Amy Connolly, Eric Huff, Tim Linden, Kenny Ng, David Nygren, Eric Oberla, Annika Peter, Benedikt Riedel, Jakob van Santen, Shigeru Yoshida, Guanying Zhu, and especially Lutz K\"opke and Anna Steuer for useful discussions and comments.  We also thank Markus Ahlers, Kfir Blum, Claudio Kopper, Carsten Rott, and others for helpful feedback beginning at J.F.B.'s presentation of these ideas at the CCAPP Cosmic Messages in Ghostly Bottles workshop in 2014.  S.W.L., M.B., and J.F.B. are supported by NSF Grant No. PHY-1404311.
\end{acknowledgments}



\begin{thebibliography}{84}%
\makeatletter
\providecommand \@ifxundefined [1]{%
 \@ifx{#1\undefined}
}%
\providecommand \@ifnum [1]{%
 \ifnum #1\expandafter \@firstoftwo
 \else \expandafter \@secondoftwo
 \fi
}%
\providecommand \@ifx [1]{%
 \ifx #1\expandafter \@firstoftwo
 \else \expandafter \@secondoftwo
 \fi
}%
\providecommand \natexlab [1]{#1}%
\providecommand \enquote  [1]{``#1''}%
\providecommand \bibnamefont  [1]{#1}%
\providecommand \bibfnamefont [1]{#1}%
\providecommand \citenamefont [1]{#1}%
\providecommand \href@noop [0]{\@secondoftwo}%
\providecommand \href [0]{\begingroup \@sanitize@url \@href}%
\providecommand \@href[1]{\@@startlink{#1}\@@href}%
\providecommand \@@href[1]{\endgroup#1\@@endlink}%
\providecommand \@sanitize@url [0]{\catcode `\\12\catcode `\$12\catcode
  `\&12\catcode `\#12\catcode `\^12\catcode `\_12\catcode `\%12\relax}%
\providecommand \@@startlink[1]{}%
\providecommand \@@endlink[0]{}%
\providecommand \url  [0]{\begingroup\@sanitize@url \@url }%
\providecommand \@url [1]{\endgroup\@href {#1}{\urlprefix }}%
\providecommand \urlprefix  [0]{URL }%
\providecommand \Eprint [0]{\href }%
\providecommand \doibase [0]{http://dx.doi.org/}%
\providecommand \selectlanguage [0]{\@gobble}%
\providecommand \bibinfo  [0]{\@secondoftwo}%
\providecommand \bibfield  [0]{\@secondoftwo}%
\providecommand \translation [1]{[#1]}%
\providecommand \BibitemOpen [0]{}%
\providecommand \bibitemStop [0]{}%
\providecommand \bibitemNoStop [0]{.\EOS\space}%
\providecommand \EOS [0]{\spacefactor3000\relax}%
\providecommand \BibitemShut  [1]{\csname bibitem#1\endcsname}%
\let\auto@bib@innerbib\@empty
\bibitem [{\citenamefont {Aartsen}\ \emph
  {et~al.}(2013{\natexlab{a}})\citenamefont {Aartsen} \emph
  {et~al.}}]{Aartsen:2013bka}%
  \BibitemOpen
  \bibfield  {author} {\bibinfo {author} {\bibfnamefont {M.~G.}\ \bibnamefont
  {Aartsen}} \emph {et~al.} (\bibinfo {collaboration} {IceCube
  Collaboration}),\ }\bibfield  {title} {\enquote {\bibinfo {title} {{First
  observation of PeV-energy neutrinos with IceCube}},}\ }\href {\doibase
  10.1103/PhysRevLett.111.021103} {\bibfield  {journal} {\bibinfo  {journal}
  {Phys.~Rev.~Lett.}\ }\textbf {\bibinfo {volume} {111}},\ \bibinfo {pages}
  {021103} (\bibinfo {year} {2013}{\natexlab{a}})},\ \Eprint
  {http://arxiv.org/abs/1304.5356} {arXiv:1304.5356 [astro-ph.HE]} \BibitemShut
  {NoStop}%
\bibitem [{\citenamefont {Aartsen}\ \emph
  {et~al.}(2013{\natexlab{b}})\citenamefont {Aartsen} \emph
  {et~al.}}]{Aartsen:2013jdh}%
  \BibitemOpen
  \bibfield  {author} {\bibinfo {author} {\bibfnamefont {M.~G.}\ \bibnamefont
  {Aartsen}} \emph {et~al.} (\bibinfo {collaboration} {IceCube}),\ }\bibfield
  {title} {\enquote {\bibinfo {title} {{Evidence for High-Energy
  Extraterrestrial Neutrinos at the IceCube Detector}},}\ }\href {\doibase
  10.1126/science.1242856} {\bibfield  {journal} {\bibinfo  {journal}
  {Science}\ }\textbf {\bibinfo {volume} {342}},\ \bibinfo {pages} {1242856}
  (\bibinfo {year} {2013}{\natexlab{b}})},\ \Eprint
  {http://arxiv.org/abs/1311.5238} {arXiv:1311.5238 [astro-ph.HE]} \BibitemShut
  {NoStop}%
\bibitem [{\citenamefont {Aartsen}\ \emph
  {et~al.}(2014{\natexlab{a}})\citenamefont {Aartsen} \emph
  {et~al.}}]{Aartsen:2013eka}%
  \BibitemOpen
  \bibfield  {author} {\bibinfo {author} {\bibfnamefont {M.~G.}\ \bibnamefont
  {Aartsen}} \emph {et~al.} (\bibinfo {collaboration} {IceCube}),\ }\bibfield
  {title} {\enquote {\bibinfo {title} {{Search for a diffuse flux of
  astrophysical muon neutrinos with the IceCube 59-string configuration}},}\
  }\href {\doibase 10.1103/PhysRevD.89.062007} {\bibfield  {journal} {\bibinfo
  {journal} {Phys.~Rev.~D}\ }\textbf {\bibinfo {volume} {89}},\ \bibinfo
  {pages} {062007} (\bibinfo {year} {2014}{\natexlab{a}})},\ \Eprint
  {http://arxiv.org/abs/1311.7048} {arXiv:1311.7048 [astro-ph.HE]} \BibitemShut
  {NoStop}%
\bibitem [{\citenamefont {Aartsen}\ \emph
  {et~al.}(2014{\natexlab{b}})\citenamefont {Aartsen} \emph
  {et~al.}}]{Aartsen:2014gkd}%
  \BibitemOpen
  \bibfield  {author} {\bibinfo {author} {\bibfnamefont {M.~G.}\ \bibnamefont
  {Aartsen}} \emph {et~al.} (\bibinfo {collaboration} {IceCube}),\ }\bibfield
  {title} {\enquote {\bibinfo {title} {{Observation of High-Energy
  Astrophysical Neutrinos in Three Years of IceCube Data}},}\ }\href {\doibase
  10.1103/PhysRevLett.113.101101} {\bibfield  {journal} {\bibinfo  {journal}
  {Phys.~Rev.~Lett.}\ }\textbf {\bibinfo {volume} {113}},\ \bibinfo {pages}
  {101101} (\bibinfo {year} {2014}{\natexlab{b}})},\ \Eprint
  {http://arxiv.org/abs/1405.5303} {arXiv:1405.5303 [astro-ph.HE]} \BibitemShut
  {NoStop}%
\bibitem [{\citenamefont {Aartsen}\ \emph
  {et~al.}(2015{\natexlab{a}})\citenamefont {Aartsen} \emph
  {et~al.}}]{Aartsen:2015ita}%
  \BibitemOpen
  \bibfield  {author} {\bibinfo {author} {\bibfnamefont {M.~G.}\ \bibnamefont
  {Aartsen}} \emph {et~al.} (\bibinfo {collaboration} {IceCube}),\ }\bibfield
  {title} {\enquote {\bibinfo {title} {{A combined maximum-likelihood analysis
  of the high-energy astrophysical neutrino flux measured with IceCube}},}\
  }\href {\doibase 10.1088/0004-637X/809/1/98} {\bibfield  {journal} {\bibinfo
  {journal} {Astrophys.~J.}\ }\textbf {\bibinfo {volume} {809}},\ \bibinfo
  {pages} {98} (\bibinfo {year} {2015}{\natexlab{a}})},\ \Eprint
  {http://arxiv.org/abs/1507.03991} {arXiv:1507.03991 [astro-ph.HE]}
  \BibitemShut {NoStop}%
\bibitem [{\citenamefont {Aartsen}\ \emph
  {et~al.}(2015{\natexlab{b}})\citenamefont {Aartsen} \emph
  {et~al.}}]{Aartsen:2015rwa}%
  \BibitemOpen
  \bibfield  {author} {\bibinfo {author} {\bibfnamefont {M.~G.}\ \bibnamefont
  {Aartsen}} \emph {et~al.} (\bibinfo {collaboration} {IceCube}),\ }\bibfield
  {title} {\enquote {\bibinfo {title} {{Evidence for Astrophysical Muon
  Neutrinos from the Northern Sky with IceCube}},}\ }\href {\doibase
  10.1103/PhysRevLett.115.081102} {\bibfield  {journal} {\bibinfo  {journal}
  {Phys. Rev. Lett.}\ }\textbf {\bibinfo {volume} {115}},\ \bibinfo {pages}
  {081102} (\bibinfo {year} {2015}{\natexlab{b}})},\ \Eprint
  {http://arxiv.org/abs/1507.04005} {arXiv:1507.04005 [astro-ph.HE]}
  \BibitemShut {NoStop}%
\bibitem [{\citenamefont {Learned}\ and\ \citenamefont
  {Pakvasa}(1995)}]{Learned:1994wg}%
  \BibitemOpen
  \bibfield  {author} {\bibinfo {author} {\bibfnamefont {J.~G.}\ \bibnamefont
  {Learned}}\ and\ \bibinfo {author} {\bibfnamefont {S.}~\bibnamefont
  {Pakvasa}},\ }\bibfield  {title} {\enquote {\bibinfo {title} {{Detecting
  tau-neutrino oscillations at PeV energies}},}\ }\href {\doibase
  10.1016/0927-6505(94)00043-3} {\bibfield  {journal} {\bibinfo  {journal}
  {Astropart.~Phys.}\ }\textbf {\bibinfo {volume} {3}},\ \bibinfo {pages}
  {267--274} (\bibinfo {year} {1995})},\ \Eprint
  {http://arxiv.org/abs/hep-ph/9405296} {arXiv:hep-ph/9405296} \BibitemShut
  {NoStop}%
\bibitem [{\citenamefont {Athar}\ \emph {et~al.}(2000)\citenamefont {Athar},
  \citenamefont {Jezabek},\ and\ \citenamefont {Yasuda}}]{Athar:2000yw}%
  \BibitemOpen
  \bibfield  {author} {\bibinfo {author} {\bibfnamefont {H.}~\bibnamefont
  {Athar}}, \bibinfo {author} {\bibfnamefont {M.}~\bibnamefont {Jezabek}}, \
  and\ \bibinfo {author} {\bibfnamefont {O.}~\bibnamefont {Yasuda}},\
  }\bibfield  {title} {\enquote {\bibinfo {title} {{Effects of neutrino mixing
  on high-energy cosmic neutrino flux}},}\ }\href {\doibase
  10.1103/PhysRevD.62.103007} {\bibfield  {journal} {\bibinfo  {journal} {Phys.
  Rev. D}\ }\textbf {\bibinfo {volume} {62}},\ \bibinfo {pages} {103007}
  (\bibinfo {year} {2000})},\ \Eprint {http://arxiv.org/abs/hep-ph/0005104}
  {arXiv:hep-ph/0005104 [hep-ph]} \BibitemShut {NoStop}%
\bibitem [{\citenamefont {Bustamante}\ \emph {et~al.}(2015)\citenamefont
  {Bustamante}, \citenamefont {Beacom},\ and\ \citenamefont
  {Winter}}]{Bustamante:2015waa}%
  \BibitemOpen
  \bibfield  {author} {\bibinfo {author} {\bibfnamefont {M.}~\bibnamefont
  {Bustamante}}, \bibinfo {author} {\bibfnamefont {J.~F.}\ \bibnamefont
  {Beacom}}, \ and\ \bibinfo {author} {\bibfnamefont {W.}~\bibnamefont
  {Winter}},\ }\bibfield  {title} {\enquote {\bibinfo {title} {{Theoretically
  palatable flavor combinations of astrophysical neutrinos}},}\ }\href
  {\doibase 10.1103/PhysRevLett.115.161302} {\bibfield  {journal} {\bibinfo
  {journal} {Phys.~Rev.~Lett.}\ }\textbf {\bibinfo {volume} {115}},\ \bibinfo
  {pages} {161302} (\bibinfo {year} {2015})},\ \Eprint
  {http://arxiv.org/abs/1506.02645} {arXiv:1506.02645 [astro-ph.HE]}
  \BibitemShut {NoStop}%
\bibitem [{\citenamefont {Beacom}\ \emph
  {et~al.}(2003{\natexlab{a}})\citenamefont {Beacom}, \citenamefont {Bell},
  \citenamefont {Hooper}, \citenamefont {Pakvasa},\ and\ \citenamefont
  {Weiler}}]{Beacom:2002vi}%
  \BibitemOpen
  \bibfield  {author} {\bibinfo {author} {\bibfnamefont {J.~F.}\ \bibnamefont
  {Beacom}}, \bibinfo {author} {\bibfnamefont {N.~F.}\ \bibnamefont {Bell}},
  \bibinfo {author} {\bibfnamefont {D.}~\bibnamefont {Hooper}}, \bibinfo
  {author} {\bibfnamefont {S.}~\bibnamefont {Pakvasa}}, \ and\ \bibinfo
  {author} {\bibfnamefont {T.~J.}\ \bibnamefont {Weiler}},\ }\bibfield  {title}
  {\enquote {\bibinfo {title} {{Decay of high-energy astrophysical
  neutrinos}},}\ }\href {\doibase 10.1103/PhysRevLett.90.181301} {\bibfield
  {journal} {\bibinfo  {journal} {Phys.~Rev.~Lett.}\ }\textbf {\bibinfo
  {volume} {90}},\ \bibinfo {pages} {181301} (\bibinfo {year}
  {2003}{\natexlab{a}})},\ \Eprint {http://arxiv.org/abs/hep-ph/0211305}
  {arXiv:hep-ph/0211305 [hep-ph]} \BibitemShut {NoStop}%
\bibitem [{\citenamefont {Barenboim}\ and\ \citenamefont
  {Quigg}(2003)}]{Barenboim:2003jm}%
  \BibitemOpen
  \bibfield  {author} {\bibinfo {author} {\bibfnamefont {G.}~\bibnamefont
  {Barenboim}}\ and\ \bibinfo {author} {\bibfnamefont {C.}~\bibnamefont
  {Quigg}},\ }\bibfield  {title} {\enquote {\bibinfo {title} {{Neutrino
  observatories can characterize cosmic sources and neutrino properties}},}\
  }\href {\doibase 10.1103/PhysRevD.67.073024} {\bibfield  {journal} {\bibinfo
  {journal} {Phys. Rev. D}\ }\textbf {\bibinfo {volume} {67}},\ \bibinfo
  {pages} {073024} (\bibinfo {year} {2003})},\ \Eprint
  {http://arxiv.org/abs/hep-ph/0301220} {arXiv:hep-ph/0301220 [hep-ph]}
  \BibitemShut {NoStop}%
\bibitem [{\citenamefont {Beacom}\ \emph
  {et~al.}(2003{\natexlab{b}})\citenamefont {Beacom}, \citenamefont {Bell},
  \citenamefont {Hooper}, \citenamefont {Pakvasa},\ and\ \citenamefont
  {Weiler}}]{Beacom:2003nh}%
  \BibitemOpen
  \bibfield  {author} {\bibinfo {author} {\bibfnamefont {J.~F.}\ \bibnamefont
  {Beacom}}, \bibinfo {author} {\bibfnamefont {N.~F.}\ \bibnamefont {Bell}},
  \bibinfo {author} {\bibfnamefont {D.}~\bibnamefont {Hooper}}, \bibinfo
  {author} {\bibfnamefont {S.}~\bibnamefont {Pakvasa}}, \ and\ \bibinfo
  {author} {\bibfnamefont {T.~J.}\ \bibnamefont {Weiler}},\ }\bibfield  {title}
  {\enquote {\bibinfo {title} {{Measuring flavor ratios of high-energy
  astrophysical neutrinos}},}\ }\href {\doibase 10.1103/PhysRevD.68.093005}
  {\bibfield  {journal} {\bibinfo  {journal} {Phys. Rev. D}\ }\textbf {\bibinfo
  {volume} {68}},\ \bibinfo {pages} {093005} (\bibinfo {year}
  {2003}{\natexlab{b}})},\ \bibinfo {note} {[Erratum: Phys. Rev. D 72, 019901
  (2005)]},\ \Eprint {http://arxiv.org/abs/hep-ph/0307025}
  {arXiv:hep-ph/0307025 [hep-ph]} \BibitemShut {NoStop}%
\bibitem [{\citenamefont {Beacom}\ \emph
  {et~al.}(2004{\natexlab{a}})\citenamefont {Beacom}, \citenamefont {Bell},
  \citenamefont {Hooper}, \citenamefont {Pakvasa},\ and\ \citenamefont
  {Weiler}}]{Beacom:2003zg}%
  \BibitemOpen
  \bibfield  {author} {\bibinfo {author} {\bibfnamefont {J.~F.}\ \bibnamefont
  {Beacom}}, \bibinfo {author} {\bibfnamefont {N.~F.}\ \bibnamefont {Bell}},
  \bibinfo {author} {\bibfnamefont {D.}~\bibnamefont {Hooper}}, \bibinfo
  {author} {\bibfnamefont {S.}~\bibnamefont {Pakvasa}}, \ and\ \bibinfo
  {author} {\bibfnamefont {T.~J.}\ \bibnamefont {Weiler}},\ }\bibfield  {title}
  {\enquote {\bibinfo {title} {{Sensitivity to $\theta_{13}$ and $\delta$ in
  the decaying astrophysical neutrino scenario}},}\ }\href {\doibase
  10.1103/PhysRevD.69.017303} {\bibfield  {journal} {\bibinfo  {journal}
  {Phys.~Rev.~D}\ }\textbf {\bibinfo {volume} {69}},\ \bibinfo {pages} {017303}
  (\bibinfo {year} {2004}{\natexlab{a}})},\ \Eprint
  {http://arxiv.org/abs/hep-ph/0309267} {arXiv:hep-ph/0309267 [hep-ph]}
  \BibitemShut {NoStop}%
\bibitem [{\citenamefont {Beacom}\ \emph
  {et~al.}(2004{\natexlab{b}})\citenamefont {Beacom}, \citenamefont {Bell},
  \citenamefont {Hooper}, \citenamefont {Learned}, \citenamefont {Pakvasa},\
  and\ \citenamefont {Weiler}}]{Beacom:2003eu}%
  \BibitemOpen
  \bibfield  {author} {\bibinfo {author} {\bibfnamefont {J.~F.}\ \bibnamefont
  {Beacom}}, \bibinfo {author} {\bibfnamefont {N.~F.}\ \bibnamefont {Bell}},
  \bibinfo {author} {\bibfnamefont {D.}~\bibnamefont {Hooper}}, \bibinfo
  {author} {\bibfnamefont {J.~G.}\ \bibnamefont {Learned}}, \bibinfo {author}
  {\bibfnamefont {S.}~\bibnamefont {Pakvasa}}, \ and\ \bibinfo {author}
  {\bibfnamefont {T.~J.}\ \bibnamefont {Weiler}},\ }\bibfield  {title}
  {\enquote {\bibinfo {title} {{Pseudo-Dirac neutrinos, a challenge for
  neutrino telescopes}},}\ }\href {\doibase 10.1103/PhysRevLett.92.011101}
  {\bibfield  {journal} {\bibinfo  {journal} {Phys. Rev. Lett.}\ }\textbf
  {\bibinfo {volume} {92}},\ \bibinfo {pages} {011101} (\bibinfo {year}
  {2004}{\natexlab{b}})},\ \Eprint {http://arxiv.org/abs/hep-ph/0307151}
  {arXiv:hep-ph/0307151} \BibitemShut {NoStop}%
\bibitem [{\citenamefont {Xing}\ and\ \citenamefont
  {Zhou}(2006)}]{Xing:2006uk}%
  \BibitemOpen
  \bibfield  {author} {\bibinfo {author} {\bibfnamefont {Z.~Z.}\ \bibnamefont
  {Xing}}\ and\ \bibinfo {author} {\bibfnamefont {S.}~\bibnamefont {Zhou}},\
  }\bibfield  {title} {\enquote {\bibinfo {title} {{Towards determination of
  the initial flavor composition of ultrahigh-energy neutrino fluxes with
  neutrino telescopes}},}\ }\href {\doibase 10.1103/PhysRevD.74.013010}
  {\bibfield  {journal} {\bibinfo  {journal} {Phys. Rev. D}\ }\textbf {\bibinfo
  {volume} {74}},\ \bibinfo {pages} {013010} (\bibinfo {year} {2006})},\
  \Eprint {http://arxiv.org/abs/astro-ph/0603781} {arXiv:astro-ph/0603781
  [astro-ph]} \BibitemShut {NoStop}%
\bibitem [{\citenamefont {Lipari}\ \emph {et~al.}(2007)\citenamefont {Lipari},
  \citenamefont {Lusignoli},\ and\ \citenamefont {Meloni}}]{Lipari:2007su}%
  \BibitemOpen
  \bibfield  {author} {\bibinfo {author} {\bibfnamefont {P.}~\bibnamefont
  {Lipari}}, \bibinfo {author} {\bibfnamefont {M.}~\bibnamefont {Lusignoli}}, \
  and\ \bibinfo {author} {\bibfnamefont {D.}~\bibnamefont {Meloni}},\
  }\bibfield  {title} {\enquote {\bibinfo {title} {{Flavor Composition and
  Energy Spectrum of Astrophysical Neutrinos}},}\ }\href {\doibase
  10.1103/PhysRevD.75.123005} {\bibfield  {journal} {\bibinfo  {journal}
  {Phys.~Rev.~D}\ }\textbf {\bibinfo {volume} {75}},\ \bibinfo {pages} {123005}
  (\bibinfo {year} {2007})},\ \Eprint {http://arxiv.org/abs/0704.0718}
  {arXiv:0704.0718 [astro-ph]} \BibitemShut {NoStop}%
\bibitem [{\citenamefont {Pakvasa}\ \emph {et~al.}(2008)\citenamefont
  {Pakvasa}, \citenamefont {Rodejohann},\ and\ \citenamefont
  {Weiler}}]{Pakvasa:2007dc}%
  \BibitemOpen
  \bibfield  {author} {\bibinfo {author} {\bibfnamefont {S.}~\bibnamefont
  {Pakvasa}}, \bibinfo {author} {\bibfnamefont {W.}~\bibnamefont {Rodejohann}},
  \ and\ \bibinfo {author} {\bibfnamefont {T.~J.}\ \bibnamefont {Weiler}},\
  }\bibfield  {title} {\enquote {\bibinfo {title} {{Flavor Ratios of
  Astrophysical Neutrinos: Implications for Precision Measurements}},}\ }\href
  {\doibase 10.1088/1126-6708/2008/02/005} {\bibfield  {journal} {\bibinfo
  {journal} {JHEP}\ }\textbf {\bibinfo {volume} {02}},\ \bibinfo {pages} {005}
  (\bibinfo {year} {2008})},\ \Eprint {http://arxiv.org/abs/0711.4517}
  {arXiv:0711.4517 [hep-ph]} \BibitemShut {NoStop}%
\bibitem [{\citenamefont {Esmaili}\ and\ \citenamefont
  {Farzan}(2009)}]{Esmaili:2009dz}%
  \BibitemOpen
  \bibfield  {author} {\bibinfo {author} {\bibfnamefont {A.}~\bibnamefont
  {Esmaili}}\ and\ \bibinfo {author} {\bibfnamefont {Y.}~\bibnamefont
  {Farzan}},\ }\bibfield  {title} {\enquote {\bibinfo {title} {{An Analysis of
  Cosmic Neutrinos: Flavor Composition at Source and Neutrino Mixing
  Parameters}},}\ }\href {\doibase 10.1016/j.nuclphysb.2009.06.017} {\bibfield
  {journal} {\bibinfo  {journal} {Nucl.~Phys.~B}\ }\textbf {\bibinfo {volume}
  {821}},\ \bibinfo {pages} {197--214} (\bibinfo {year} {2009})},\ \Eprint
  {http://arxiv.org/abs/0905.0259} {arXiv:0905.0259 [hep-ph]} \BibitemShut
  {NoStop}%
\bibitem [{\citenamefont {Lai}\ \emph {et~al.}(2009)\citenamefont {Lai},
  \citenamefont {Lin},\ and\ \citenamefont {Liu}}]{Lai:2009ke}%
  \BibitemOpen
  \bibfield  {author} {\bibinfo {author} {\bibfnamefont {K.~C.}\ \bibnamefont
  {Lai}}, \bibinfo {author} {\bibfnamefont {G.~L.}\ \bibnamefont {Lin}}, \ and\
  \bibinfo {author} {\bibfnamefont {T.~C.}\ \bibnamefont {Liu}},\ }\bibfield
  {title} {\enquote {\bibinfo {title} {{Determination of the Neutrino Flavor
  Ratio at the Astrophysical Source}},}\ }\href {\doibase
  10.1103/PhysRevD.80.103005} {\bibfield  {journal} {\bibinfo  {journal}
  {Phys.~Rev.~D}\ }\textbf {\bibinfo {volume} {80}},\ \bibinfo {pages} {103005}
  (\bibinfo {year} {2009})},\ \Eprint {http://arxiv.org/abs/0905.4003}
  {arXiv:0905.4003 [hep-ph]} \BibitemShut {NoStop}%
\bibitem [{\citenamefont {Bazo}\ \emph {et~al.}(2009)\citenamefont {Bazo},
  \citenamefont {Bustamante}, \citenamefont {Gago},\ and\ \citenamefont
  {Miranda}}]{Bazo:2009en}%
  \BibitemOpen
  \bibfield  {author} {\bibinfo {author} {\bibfnamefont {J.~L.}\ \bibnamefont
  {Bazo}}, \bibinfo {author} {\bibfnamefont {M.}~\bibnamefont {Bustamante}},
  \bibinfo {author} {\bibfnamefont {A.~M.}\ \bibnamefont {Gago}}, \ and\
  \bibinfo {author} {\bibfnamefont {O.~G.}\ \bibnamefont {Miranda}},\
  }\bibfield  {title} {\enquote {\bibinfo {title} {{High energy astrophysical
  neutrino flux and modified dispersion relations}},}\ }\href {\doibase
  10.1142/S0217751X09047429} {\bibfield  {journal} {\bibinfo  {journal} {Int.\
  J.\ Mod.\ Phys.\ A}\ }\textbf {\bibinfo {volume} {24}},\ \bibinfo {pages}
  {5819} (\bibinfo {year} {2009})},\ \Eprint {http://arxiv.org/abs/0907.1979}
  {arXiv:0907.1979 [hep-ph]} \BibitemShut {NoStop}%
\bibitem [{\citenamefont {Choubey}\ and\ \citenamefont
  {Rodejohann}(2009)}]{Choubey:2009jq}%
  \BibitemOpen
  \bibfield  {author} {\bibinfo {author} {\bibfnamefont {S.}~\bibnamefont
  {Choubey}}\ and\ \bibinfo {author} {\bibfnamefont {W.}~\bibnamefont
  {Rodejohann}},\ }\bibfield  {title} {\enquote {\bibinfo {title} {{Flavor
  Composition of UHE Neutrinos at Source and at Neutrino Telescopes}},}\ }\href
  {\doibase 10.1103/PhysRevD.80.113006} {\bibfield  {journal} {\bibinfo
  {journal} {Phys.~Rev.~D}\ }\textbf {\bibinfo {volume} {80}},\ \bibinfo
  {pages} {113006} (\bibinfo {year} {2009})},\ \Eprint
  {http://arxiv.org/abs/0909.1219} {arXiv:0909.1219 [hep-ph]} \BibitemShut
  {NoStop}%
\bibitem [{\citenamefont {Bustamante}\ \emph {et~al.}(2010)\citenamefont
  {Bustamante}, \citenamefont {Gago},\ and\ \citenamefont
  {Pena-Garay}}]{Bustamante:2010nq}%
  \BibitemOpen
  \bibfield  {author} {\bibinfo {author} {\bibfnamefont {M.}~\bibnamefont
  {Bustamante}}, \bibinfo {author} {\bibfnamefont {A.~M.}\ \bibnamefont
  {Gago}}, \ and\ \bibinfo {author} {\bibfnamefont {C.}~\bibnamefont
  {Pena-Garay}},\ }\bibfield  {title} {\enquote {\bibinfo {title}
  {{Energy-independent new physics in the flavour ratios of high-energy
  astrophysical neutrinos}},}\ }\href {\doibase 10.1007/JHEP04(2010)066}
  {\bibfield  {journal} {\bibinfo  {journal} {JHEP}\ }\textbf {\bibinfo
  {volume} {1004}},\ \bibinfo {pages} {066} (\bibinfo {year} {2010})},\ \Eprint
  {http://arxiv.org/abs/1001.4878} {arXiv:1001.4878 [hep-ph]} \BibitemShut
  {NoStop}%
\bibitem [{\citenamefont {Bustamante}\ \emph {et~al.}(2011)\citenamefont
  {Bustamante}, \citenamefont {Gago},\ and\ \citenamefont
  {Jones-Perez}}]{Bustamante:2010bf}%
  \BibitemOpen
  \bibfield  {author} {\bibinfo {author} {\bibfnamefont {M.}~\bibnamefont
  {Bustamante}}, \bibinfo {author} {\bibfnamefont {A.~M.}\ \bibnamefont
  {Gago}}, \ and\ \bibinfo {author} {\bibfnamefont {J.}~\bibnamefont
  {Jones-Perez}},\ }\bibfield  {title} {\enquote {\bibinfo {title} {{SUSY
  Renormalization Group Effects in Ultra High Energy Neutrinos}},}\ }\href
  {\doibase 10.1007/JHEP05(2011)133} {\bibfield  {journal} {\bibinfo  {journal}
  {JHEP}\ }\textbf {\bibinfo {volume} {1105}},\ \bibinfo {pages} {133}
  (\bibinfo {year} {2011})},\ \Eprint {http://arxiv.org/abs/1012.2728}
  {arXiv:1012.2728 [hep-ph]} \BibitemShut {NoStop}%
\bibitem [{\citenamefont {Baerwald}\ \emph {et~al.}(2012)\citenamefont
  {Baerwald}, \citenamefont {Bustamante},\ and\ \citenamefont
  {Winter}}]{Baerwald:2012kc}%
  \BibitemOpen
  \bibfield  {author} {\bibinfo {author} {\bibfnamefont {P.}~\bibnamefont
  {Baerwald}}, \bibinfo {author} {\bibfnamefont {M.}~\bibnamefont
  {Bustamante}}, \ and\ \bibinfo {author} {\bibfnamefont {W.}~\bibnamefont
  {Winter}},\ }\bibfield  {title} {\enquote {\bibinfo {title} {{Neutrino Decays
  over Cosmological Distances and the Implications for Neutrino Telescopes}},}\
  }\href {\doibase 10.1088/1475-7516/2012/10/020} {\bibfield  {journal}
  {\bibinfo  {journal} {JCAP}\ }\textbf {\bibinfo {volume} {1210}},\ \bibinfo
  {pages} {020} (\bibinfo {year} {2012})},\ \Eprint
  {http://arxiv.org/abs/1208.4600} {arXiv:1208.4600 [astro-ph.CO]} \BibitemShut
  {NoStop}%
\bibitem [{\citenamefont {Laha}\ \emph {et~al.}(2013)\citenamefont {Laha},
  \citenamefont {Beacom}, \citenamefont {Dasgupta}, \citenamefont {Horiuchi},\
  and\ \citenamefont {Murase}}]{Laha:2013lka}%
  \BibitemOpen
  \bibfield  {author} {\bibinfo {author} {\bibfnamefont {R.}~\bibnamefont
  {Laha}}, \bibinfo {author} {\bibfnamefont {J.~F.}\ \bibnamefont {Beacom}},
  \bibinfo {author} {\bibfnamefont {B.}~\bibnamefont {Dasgupta}}, \bibinfo
  {author} {\bibfnamefont {S.}~\bibnamefont {Horiuchi}}, \ and\ \bibinfo
  {author} {\bibfnamefont {K.}~\bibnamefont {Murase}},\ }\bibfield  {title}
  {\enquote {\bibinfo {title} {{Demystifying the PeV Cascades in IceCube: Less
  (Energy) is More (Events)}},}\ }\href {\doibase 10.1103/PhysRevD.88.043009}
  {\bibfield  {journal} {\bibinfo  {journal} {Phys.~Rev.~D}\ }\textbf {\bibinfo
  {volume} {88}},\ \bibinfo {pages} {043009} (\bibinfo {year} {2013})},\
  \Eprint {http://arxiv.org/abs/1306.2309} {arXiv:1306.2309 [astro-ph.HE]}
  \BibitemShut {NoStop}%
\bibitem [{\citenamefont {Chen}\ \emph {et~al.}(2014)\citenamefont {Chen},
  \citenamefont {Dev},\ and\ \citenamefont {Soni}}]{Chen:2013dza}%
  \BibitemOpen
  \bibfield  {author} {\bibinfo {author} {\bibfnamefont {C.~Y.}\ \bibnamefont
  {Chen}}, \bibinfo {author} {\bibfnamefont {P.~S.~Bhupal}\ \bibnamefont
  {Dev}}, \ and\ \bibinfo {author} {\bibfnamefont {A.}~\bibnamefont {Soni}},\
  }\bibfield  {title} {\enquote {\bibinfo {title} {{Standard model explanation
  of the ultrahigh energy neutrino events at IceCube}},}\ }\href {\doibase
  10.1103/PhysRevD.89.033012} {\bibfield  {journal} {\bibinfo  {journal} {Phys.
  Rev. D}\ }\textbf {\bibinfo {volume} {89}},\ \bibinfo {pages} {033012}
  (\bibinfo {year} {2014})},\ \Eprint {http://arxiv.org/abs/1309.1764}
  {arXiv:1309.1764 [hep-ph]} \BibitemShut {NoStop}%
\bibitem [{\citenamefont {Mena}\ \emph {et~al.}(2014)\citenamefont {Mena},
  \citenamefont {Palomares-Ruiz},\ and\ \citenamefont
  {Vincent}}]{Mena:2014sja}%
  \BibitemOpen
  \bibfield  {author} {\bibinfo {author} {\bibfnamefont {O.}~\bibnamefont
  {Mena}}, \bibinfo {author} {\bibfnamefont {S.}~\bibnamefont
  {Palomares-Ruiz}}, \ and\ \bibinfo {author} {\bibfnamefont {A.~C.}\
  \bibnamefont {Vincent}},\ }\bibfield  {title} {\enquote {\bibinfo {title}
  {{Flavor Composition of the High-Energy Neutrino Events in IceCube}},}\
  }\href {\doibase 10.1103/PhysRevLett.113.091103} {\bibfield  {journal}
  {\bibinfo  {journal} {Phys.~Rev.~Lett.}\ }\textbf {\bibinfo {volume} {113}},\
  \bibinfo {pages} {091103} (\bibinfo {year} {2014})},\ \Eprint
  {http://arxiv.org/abs/1404.0017} {arXiv:1404.0017 [astro-ph.HE]} \BibitemShut
  {NoStop}%
\bibitem [{\citenamefont {Ng}\ and\ \citenamefont {Beacom}(2014)}]{Ng:2014pca}%
  \BibitemOpen
  \bibfield  {author} {\bibinfo {author} {\bibfnamefont {K.~C.~Y.}\
  \bibnamefont {Ng}}\ and\ \bibinfo {author} {\bibfnamefont {J.~F.}\
  \bibnamefont {Beacom}},\ }\bibfield  {title} {\enquote {\bibinfo {title}
  {{Cosmic neutrino cascades from secret neutrino interactions}},}\ }\href
  {\doibase 10.1103/PhysRevD.90.065035, 10.1103/PhysRevD.90.089904} {\bibfield
  {journal} {\bibinfo  {journal} {Phys.\ Rev.\ D}\ }\textbf {\bibinfo {volume}
  {90}},\ \bibinfo {pages} {065035} (\bibinfo {year} {2014})},\ \Eprint
  {http://arxiv.org/abs/1404.2288} {arXiv:1404.2288 [astro-ph.HE]} \BibitemShut
  {NoStop}%
\bibitem [{\citenamefont {Xu}\ \emph {et~al.}(2014)\citenamefont {Xu},
  \citenamefont {He},\ and\ \citenamefont {Rodejohann}}]{Xu:2014via}%
  \BibitemOpen
  \bibfield  {author} {\bibinfo {author} {\bibfnamefont {X.}~\bibnamefont
  {Xu}}, \bibinfo {author} {\bibfnamefont {H.}~\bibnamefont {He}}, \ and\
  \bibinfo {author} {\bibfnamefont {W.}~\bibnamefont {Rodejohann}},\ }\bibfield
   {title} {\enquote {\bibinfo {title} {{Constraining Astrophysical Neutrino
  Flavor Composition from Leptonic Unitarity}},}\ }\href {\doibase
  10.1088/1475-7516/2014/12/039} {\bibfield  {journal} {\bibinfo  {journal}
  {JCAP}\ }\textbf {\bibinfo {volume} {1412}},\ \bibinfo {pages} {039}
  (\bibinfo {year} {2014})},\ \Eprint {http://arxiv.org/abs/1407.3736}
  {arXiv:1407.3736 [hep-ph]} \BibitemShut {NoStop}%
\bibitem [{\citenamefont {Fu}\ \emph {et~al.}(2015)\citenamefont {Fu},
  \citenamefont {Ho},\ and\ \citenamefont {Weiler}}]{Fu:2014isa}%
  \BibitemOpen
  \bibfield  {author} {\bibinfo {author} {\bibfnamefont {L.}~\bibnamefont
  {Fu}}, \bibinfo {author} {\bibfnamefont {C.~M.}\ \bibnamefont {Ho}}, \ and\
  \bibinfo {author} {\bibfnamefont {T.~J.}\ \bibnamefont {Weiler}},\ }\bibfield
   {title} {\enquote {\bibinfo {title} {{Aspects of the Flavor Triangle for
  Cosmic Neutrino Propagation}},}\ }\href {\doibase 10.1103/PhysRevD.91.053001}
  {\bibfield  {journal} {\bibinfo  {journal} {Phys.~Rev.~D}\ }\textbf {\bibinfo
  {volume} {91}},\ \bibinfo {pages} {053001} (\bibinfo {year} {2015})},\
  \Eprint {http://arxiv.org/abs/1411.1174} {arXiv:1411.1174 [hep-ph]}
  \BibitemShut {NoStop}%
\bibitem [{\citenamefont {Chen}\ \emph {et~al.}(2015)\citenamefont {Chen},
  \citenamefont {Dev},\ and\ \citenamefont {Soni}}]{Chen:2014gxa}%
  \BibitemOpen
  \bibfield  {author} {\bibinfo {author} {\bibfnamefont {C.~Y.}\ \bibnamefont
  {Chen}}, \bibinfo {author} {\bibfnamefont {P.~S.~Bhupal}\ \bibnamefont
  {Dev}}, \ and\ \bibinfo {author} {\bibfnamefont {A.}~\bibnamefont {Soni}},\
  }\bibfield  {title} {\enquote {\bibinfo {title} {{Two-component flux
  explanation for the high energy neutrino events at IceCube}},}\ }\href
  {\doibase 10.1103/PhysRevD.92.073001} {\bibfield  {journal} {\bibinfo
  {journal} {Phys. Rev. D}\ }\textbf {\bibinfo {volume} {92}},\ \bibinfo
  {pages} {073001} (\bibinfo {year} {2015})},\ \Eprint
  {http://arxiv.org/abs/1411.5658} {arXiv:1411.5658 [hep-ph]} \BibitemShut
  {NoStop}%
\bibitem [{\citenamefont {Palomares-Ruiz}\ \emph {et~al.}(2015)\citenamefont
  {Palomares-Ruiz}, \citenamefont {Vincent},\ and\ \citenamefont
  {Mena}}]{Palomares-Ruiz:2015mka}%
  \BibitemOpen
  \bibfield  {author} {\bibinfo {author} {\bibfnamefont {S.}~\bibnamefont
  {Palomares-Ruiz}}, \bibinfo {author} {\bibfnamefont {A.~C.}\ \bibnamefont
  {Vincent}}, \ and\ \bibinfo {author} {\bibfnamefont {O.}~\bibnamefont
  {Mena}},\ }\bibfield  {title} {\enquote {\bibinfo {title} {{Spectral analysis
  of the high-energy IceCube neutrinos}},}\ }\href {\doibase
  10.1103/PhysRevD.91.103008} {\bibfield  {journal} {\bibinfo  {journal}
  {Phys.~Rev.~D}\ }\textbf {\bibinfo {volume} {91}},\ \bibinfo {pages} {103008}
  (\bibinfo {year} {2015})},\ \Eprint {http://arxiv.org/abs/1502.02649}
  {arXiv:1502.02649 [astro-ph.HE]} \BibitemShut {NoStop}%
\bibitem [{\citenamefont {Palladino}\ \emph {et~al.}(2015)\citenamefont
  {Palladino}, \citenamefont {Pagliaroli}, \citenamefont {Villante},\ and\
  \citenamefont {Vissani}}]{Palladino:2015zua}%
  \BibitemOpen
  \bibfield  {author} {\bibinfo {author} {\bibfnamefont {A.}~\bibnamefont
  {Palladino}}, \bibinfo {author} {\bibfnamefont {G.}~\bibnamefont
  {Pagliaroli}}, \bibinfo {author} {\bibfnamefont {F.~L.}\ \bibnamefont
  {Villante}}, \ and\ \bibinfo {author} {\bibfnamefont {F.}~\bibnamefont
  {Vissani}},\ }\bibfield  {title} {\enquote {\bibinfo {title} {{What is the
  Flavor of the Cosmic Neutrinos Seen by IceCube?}}}\ }\href {\doibase
  10.1103/PhysRevLett.114.171101} {\bibfield  {journal} {\bibinfo  {journal}
  {Phys.~Rev.~Lett.}\ }\textbf {\bibinfo {volume} {114}},\ \bibinfo {pages}
  {171101} (\bibinfo {year} {2015})},\ \Eprint
  {http://arxiv.org/abs/1502.02923} {arXiv:1502.02923 [astro-ph.HE]}
  \BibitemShut {NoStop}%
\bibitem [{\citenamefont {Aartsen}\ \emph
  {et~al.}(2015{\natexlab{c}})\citenamefont {Aartsen} \emph
  {et~al.}}]{Aartsen:2015ivb}%
  \BibitemOpen
  \bibfield  {author} {\bibinfo {author} {\bibfnamefont {M.~G.}\ \bibnamefont
  {Aartsen}} \emph {et~al.} (\bibinfo {collaboration} {IceCube}),\ }\bibfield
  {title} {\enquote {\bibinfo {title} {{Flavor Ratio of Astrophysical Neutrinos
  above 35 TeV in IceCube}},}\ }\href {\doibase 10.1103/PhysRevLett.114.171102}
  {\bibfield  {journal} {\bibinfo  {journal} {Phys.~Rev.~Lett.}\ }\textbf
  {\bibinfo {volume} {114}},\ \bibinfo {pages} {171102} (\bibinfo {year}
  {2015}{\natexlab{c}})},\ \Eprint {http://arxiv.org/abs/1502.03376}
  {arXiv:1502.03376 [astro-ph.HE]} \BibitemShut {NoStop}%
\bibitem [{\citenamefont {Palladino}\ and\ \citenamefont
  {Vissani}(2015)}]{Palladino:2015vna}%
  \BibitemOpen
  \bibfield  {author} {\bibinfo {author} {\bibfnamefont {A.}~\bibnamefont
  {Palladino}}\ and\ \bibinfo {author} {\bibfnamefont {F.}~\bibnamefont
  {Vissani}},\ }\bibfield  {title} {\enquote {\bibinfo {title} {{The natural
  parameterization of cosmic neutrino oscillations}},}\ }\href {\doibase
  10.1140/epjc/s10052-015-3664-6} {\bibfield  {journal} {\bibinfo  {journal}
  {Eur.~Phys.~J.~C}\ }\textbf {\bibinfo {volume} {75}},\ \bibinfo {pages} {433}
  (\bibinfo {year} {2015})},\ \Eprint {http://arxiv.org/abs/1504.05238}
  {arXiv:1504.05238 [hep-ph]} \BibitemShut {NoStop}%
\bibitem [{\citenamefont {Arguelles}\ \emph {et~al.}(2015)\citenamefont
  {Arguelles}, \citenamefont {Katori},\ and\ \citenamefont
  {Salvado}}]{Arguelles:2015dca}%
  \BibitemOpen
  \bibfield  {author} {\bibinfo {author} {\bibfnamefont {C.~A.}\ \bibnamefont
  {Arguelles}}, \bibinfo {author} {\bibfnamefont {T.}~\bibnamefont {Katori}}, \
  and\ \bibinfo {author} {\bibfnamefont {J.}~\bibnamefont {Salvado}},\
  }\bibfield  {title} {\enquote {\bibinfo {title} {{New Physics in
  Astrophysical Neutrino Flavor}},}\ }\href {\doibase
  10.1103/PhysRevLett.115.161303} {\bibfield  {journal} {\bibinfo  {journal}
  {Phys. Rev. Lett.}\ }\textbf {\bibinfo {volume} {115}},\ \bibinfo {pages}
  {161303} (\bibinfo {year} {2015})},\ \Eprint
  {http://arxiv.org/abs/1506.02043} {arXiv:1506.02043 [hep-ph]} \BibitemShut
  {NoStop}%
\bibitem [{\citenamefont {Palladino}\ \emph {et~al.}(2016)\citenamefont
  {Palladino}, \citenamefont {Pagliaroli}, \citenamefont {Villante},\ and\
  \citenamefont {Vissani}}]{Palladino:2015uoa}%
  \BibitemOpen
  \bibfield  {author} {\bibinfo {author} {\bibfnamefont {A.}~\bibnamefont
  {Palladino}}, \bibinfo {author} {\bibfnamefont {G.}~\bibnamefont
  {Pagliaroli}}, \bibinfo {author} {\bibfnamefont {F.~L.}\ \bibnamefont
  {Villante}}, \ and\ \bibinfo {author} {\bibfnamefont {F.}~\bibnamefont
  {Vissani}},\ }\bibfield  {title} {\enquote {\bibinfo {title} {{Double pulses
  and cascades above 2 PeV in IceCube}},}\ }\href {\doibase
  10.1140/epjc/s10052-016-3893-3} {\bibfield  {journal} {\bibinfo  {journal}
  {Eur. Phys. J. C}\ }\textbf {\bibinfo {volume} {76}},\ \bibinfo {pages} {52}
  (\bibinfo {year} {2016})},\ \Eprint {http://arxiv.org/abs/1510.05921}
  {arXiv:1510.05921 [astro-ph.HE]} \BibitemShut {NoStop}%
\bibitem [{\citenamefont {Shoemaker}\ and\ \citenamefont
  {Murase}(2016)}]{Shoemaker:2015qul}%
  \BibitemOpen
  \bibfield  {author} {\bibinfo {author} {\bibfnamefont {I.~M.}\ \bibnamefont
  {Shoemaker}}\ and\ \bibinfo {author} {\bibfnamefont {K.}~\bibnamefont
  {Murase}},\ }\bibfield  {title} {\enquote {\bibinfo {title} {{Probing BSM
  Neutrino Physics with Flavor and Spectral Distortions: Prospects for Future
  High-Energy Neutrino Telescopes}},}\ }\href {\doibase
  10.1103/PhysRevD.93.085004} {\bibfield  {journal} {\bibinfo  {journal}
  {Phys.\ Rev.\ D}\ }\textbf {\bibinfo {volume} {93}},\ \bibinfo {pages}
  {085004} (\bibinfo {year} {2016})},\ \Eprint
  {http://arxiv.org/abs/1512.07228} {arXiv:1512.07228 [astro-ph.HE]}
  \BibitemShut {NoStop}%
\bibitem [{\citenamefont {Vincent}\ \emph {et~al.}(2016)\citenamefont
  {Vincent}, \citenamefont {Palomares-Ruiz},\ and\ \citenamefont
  {Mena}}]{Vincent:2016nut}%
  \BibitemOpen
  \bibfield  {author} {\bibinfo {author} {\bibfnamefont {Aaron~C.}\
  \bibnamefont {Vincent}}, \bibinfo {author} {\bibfnamefont {Sergio}\
  \bibnamefont {Palomares-Ruiz}}, \ and\ \bibinfo {author} {\bibfnamefont
  {Olga}\ \bibnamefont {Mena}},\ }\bibfield  {title} {\enquote {\bibinfo
  {title} {{Analysis of the 4-year IceCube high-energy starting events}},}\
  }\href {\doibase 10.1103/PhysRevD.94.023009} {\bibfield  {journal} {\bibinfo
  {journal} {Phys. Rev. D}\ }\textbf {\bibinfo {volume} {94}},\ \bibinfo
  {pages} {023009} (\bibinfo {year} {2016})},\ \Eprint
  {http://arxiv.org/abs/1605.01556} {arXiv:1605.01556 [astro-ph.HE]}
  \BibitemShut {NoStop}%
\bibitem [{\citenamefont {Glashow}(1960)}]{Glashow:1960zz}%
  \BibitemOpen
  \bibfield  {author} {\bibinfo {author} {\bibfnamefont {S.~L.}\ \bibnamefont
  {Glashow}},\ }\bibfield  {title} {\enquote {\bibinfo {title} {{Resonant
  Scattering of Antineutrinos}},}\ }\href {\doibase 10.1103/PhysRev.118.316}
  {\bibfield  {journal} {\bibinfo  {journal} {Phys.~Rev.}\ }\textbf {\bibinfo
  {volume} {118}},\ \bibinfo {pages} {316} (\bibinfo {year}
  {1960})}\BibitemShut {NoStop}%
\bibitem [{\citenamefont {Anchordoqui}\ \emph {et~al.}(2005)\citenamefont
  {Anchordoqui}, \citenamefont {Goldberg}, \citenamefont {Halzen},\ and\
  \citenamefont {Weiler}}]{Anchordoqui:2004eb}%
  \BibitemOpen
  \bibfield  {author} {\bibinfo {author} {\bibfnamefont {L.~A.}\ \bibnamefont
  {Anchordoqui}}, \bibinfo {author} {\bibfnamefont {H.}~\bibnamefont
  {Goldberg}}, \bibinfo {author} {\bibfnamefont {F.}~\bibnamefont {Halzen}}, \
  and\ \bibinfo {author} {\bibfnamefont {T.~J.}\ \bibnamefont {Weiler}},\
  }\bibfield  {title} {\enquote {\bibinfo {title} {{Neutrinos as a diagnostic
  of high energy astrophysical processes}},}\ }\href {\doibase
  10.1016/j.physletb.2005.06.056} {\bibfield  {journal} {\bibinfo  {journal}
  {Phys.~Lett.~B}\ }\textbf {\bibinfo {volume} {621}},\ \bibinfo {pages}
  {18--21} (\bibinfo {year} {2005})},\ \Eprint
  {http://arxiv.org/abs/hep-ph/0410003} {arXiv:hep-ph/0410003 [hep-ph]}
  \BibitemShut {NoStop}%
\bibitem [{\citenamefont {Bhattacharya}\ \emph {et~al.}(2011)\citenamefont
  {Bhattacharya}, \citenamefont {Gandhi}, \citenamefont {Rodejohann},\ and\
  \citenamefont {Watanabe}}]{Bhattacharya:2011qu}%
  \BibitemOpen
  \bibfield  {author} {\bibinfo {author} {\bibfnamefont {A.}~\bibnamefont
  {Bhattacharya}}, \bibinfo {author} {\bibfnamefont {R.}~\bibnamefont
  {Gandhi}}, \bibinfo {author} {\bibfnamefont {W.}~\bibnamefont {Rodejohann}},
  \ and\ \bibinfo {author} {\bibfnamefont {A.}~\bibnamefont {Watanabe}},\
  }\bibfield  {title} {\enquote {\bibinfo {title} {{The Glashow resonance at
  IceCube: signatures, event rates and $pp$ vs. $p\gamma$ interactions}},}\
  }\href {\doibase 10.1088/1475-7516/2011/10/017} {\bibfield  {journal}
  {\bibinfo  {journal} {JCAP}\ }\textbf {\bibinfo {volume} {1110}},\ \bibinfo
  {pages} {017} (\bibinfo {year} {2011})},\ \Eprint
  {http://arxiv.org/abs/1108.3163} {arXiv:1108.3163 [astro-ph.HE]} \BibitemShut
  {NoStop}%
\bibitem [{\citenamefont {Aartsen}\ \emph {et~al.}(2016)\citenamefont {Aartsen}
  \emph {et~al.}}]{Aartsen:2015dlt}%
  \BibitemOpen
  \bibfield  {author} {\bibinfo {author} {\bibfnamefont {M.~G.}\ \bibnamefont
  {Aartsen}} \emph {et~al.} (\bibinfo {collaboration} {IceCube}),\ }\bibfield
  {title} {\enquote {\bibinfo {title} {{Search for Astrophysical Tau Neutrinos
  in Three Years of IceCube Data}},}\ }\href {\doibase
  10.1103/PhysRevD.93.022001} {\bibfield  {journal} {\bibinfo  {journal}
  {Phys.\ Rev.\ D}\ }\textbf {\bibinfo {volume} {93}},\ \bibinfo {pages}
  {022001} (\bibinfo {year} {2016})},\ \Eprint
  {http://arxiv.org/abs/1509.06212} {arXiv:1509.06212 [astro-ph.HE]}
  \BibitemShut {NoStop}%
\bibitem [{\citenamefont {DeYoung}\ \emph {et~al.}(2007)\citenamefont
  {DeYoung}, \citenamefont {Razzaque},\ and\ \citenamefont
  {Cowen}}]{DeYoung:2006fg}%
  \BibitemOpen
  \bibfield  {author} {\bibinfo {author} {\bibfnamefont {T.}~\bibnamefont
  {DeYoung}}, \bibinfo {author} {\bibfnamefont {S.}~\bibnamefont {Razzaque}}, \
  and\ \bibinfo {author} {\bibfnamefont {D.~F.}\ \bibnamefont {Cowen}},\
  }\bibfield  {title} {\enquote {\bibinfo {title} {{Astrophysical tau neutrino
  detection in kilometer-scale Cherenkov detectors via muonic tau decay}},}\
  }\href {\doibase 10.1016/j.astropartphys.2006.11.003} {\bibfield  {journal}
  {\bibinfo  {journal} {Astropart.~Phys.}\ }\textbf {\bibinfo {volume} {27}},\
  \bibinfo {pages} {238--243} (\bibinfo {year} {2007})},\ \Eprint
  {http://arxiv.org/abs/astro-ph/0608486} {arXiv:astro-ph/0608486 [astro-ph]}
  \BibitemShut {NoStop}%
\bibitem [{\citenamefont {Gandhi}\ \emph {et~al.}(1996)\citenamefont {Gandhi},
  \citenamefont {Quigg}, \citenamefont {Reno},\ and\ \citenamefont
  {Sarcevic}}]{Gandhi:1995tf}%
  \BibitemOpen
  \bibfield  {author} {\bibinfo {author} {\bibfnamefont {R.}~\bibnamefont
  {Gandhi}}, \bibinfo {author} {\bibfnamefont {C.}~\bibnamefont {Quigg}},
  \bibinfo {author} {\bibfnamefont {M.~H.}\ \bibnamefont {Reno}}, \ and\
  \bibinfo {author} {\bibfnamefont {I.}~\bibnamefont {Sarcevic}},\ }\bibfield
  {title} {\enquote {\bibinfo {title} {{Ultrahigh-energy neutrino
  interactions}},}\ }\href {\doibase 10.1016/0927-6505(96)00008-4} {\bibfield
  {journal} {\bibinfo  {journal} {Astropart. Phys.}\ }\textbf {\bibinfo
  {volume} {5}},\ \bibinfo {pages} {81--110} (\bibinfo {year} {1996})},\
  \Eprint {http://arxiv.org/abs/hep-ph/9512364} {arXiv:hep-ph/9512364 [hep-ph]}
  \BibitemShut {NoStop}%
\bibitem [{\citenamefont {Gandhi}\ \emph {et~al.}(1998)\citenamefont {Gandhi},
  \citenamefont {Quigg}, \citenamefont {Reno},\ and\ \citenamefont
  {Sarcevic}}]{Gandhi:1998ri}%
  \BibitemOpen
  \bibfield  {author} {\bibinfo {author} {\bibfnamefont {R.}~\bibnamefont
  {Gandhi}}, \bibinfo {author} {\bibfnamefont {C.}~\bibnamefont {Quigg}},
  \bibinfo {author} {\bibfnamefont {M.~H.}\ \bibnamefont {Reno}}, \ and\
  \bibinfo {author} {\bibfnamefont {I.}~\bibnamefont {Sarcevic}},\ }\bibfield
  {title} {\enquote {\bibinfo {title} {{Neutrino interactions at
  ultrahigh-energies}},}\ }\href {\doibase 10.1103/PhysRevD.58.093009}
  {\bibfield  {journal} {\bibinfo  {journal} {Phys. Rev. D}\ }\textbf {\bibinfo
  {volume} {58}},\ \bibinfo {pages} {093009} (\bibinfo {year} {1998})},\
  \Eprint {http://arxiv.org/abs/hep-ph/9807264} {arXiv:hep-ph/9807264 [hep-ph]}
  \BibitemShut {NoStop}%
\bibitem [{\citenamefont {Connolly}\ \emph {et~al.}(2011)\citenamefont
  {Connolly}, \citenamefont {Thorne},\ and\ \citenamefont
  {Waters}}]{Connolly:2011vc}%
  \BibitemOpen
  \bibfield  {author} {\bibinfo {author} {\bibfnamefont {A.}~\bibnamefont
  {Connolly}}, \bibinfo {author} {\bibfnamefont {R.~S.}\ \bibnamefont
  {Thorne}}, \ and\ \bibinfo {author} {\bibfnamefont {D.}~\bibnamefont
  {Waters}},\ }\bibfield  {title} {\enquote {\bibinfo {title} {{Calculation of
  High Energy Neutrino-Nucleon Cross Sections and Uncertainties Using the MSTW
  Parton Distribution Functions and Implications for Future Experiments}},}\
  }\href {\doibase 10.1103/PhysRevD.83.113009} {\bibfield  {journal} {\bibinfo
  {journal} {Phys.~Rev.~D}\ }\textbf {\bibinfo {volume} {83}},\ \bibinfo
  {pages} {113009} (\bibinfo {year} {2011})},\ \Eprint
  {http://arxiv.org/abs/1102.0691} {arXiv:1102.0691 [hep-ph]} \BibitemShut
  {NoStop}%
\bibitem [{\citenamefont {Olive}\ \emph {et~al.}(2014)\citenamefont {Olive}
  \emph {et~al.}}]{Agashe:2014kda}%
  \BibitemOpen
  \bibfield  {author} {\bibinfo {author} {\bibfnamefont {K.~A.}\ \bibnamefont
  {Olive}} \emph {et~al.} (\bibinfo {collaboration} {Particle Data Group}),\
  }\bibfield  {title} {\enquote {\bibinfo {title} {{Review of Particle
  Physics}},}\ }\href {\doibase 10.1088/1674-1137/38/9/090001} {\bibfield
  {journal} {\bibinfo  {journal} {Chin.\ Phys.\ C}\ }\textbf {\bibinfo {volume}
  {38}},\ \bibinfo {pages} {090001} (\bibinfo {year} {2014})}\BibitemShut
  {NoStop}%
\bibitem [{\citenamefont {Beacom}\ and\ \citenamefont
  {Candia}(2004)}]{Beacom:2004jb}%
  \BibitemOpen
  \bibfield  {author} {\bibinfo {author} {\bibfnamefont {J.~F.}\ \bibnamefont
  {Beacom}}\ and\ \bibinfo {author} {\bibfnamefont {J.}~\bibnamefont
  {Candia}},\ }\bibfield  {title} {\enquote {\bibinfo {title} {{Shower power:
  Isolating the prompt atmospheric neutrino flux using electron neutrinos}},}\
  }\href {\doibase 10.1088/1475-7516/2004/11/009} {\bibfield  {journal}
  {\bibinfo  {journal} {JCAP}\ }\textbf {\bibinfo {volume} {0411}},\ \bibinfo
  {pages} {009} (\bibinfo {year} {2004})},\ \Eprint
  {http://arxiv.org/abs/hep-ph/0409046} {arXiv:hep-ph/0409046 [hep-ph]}
  \BibitemShut {NoStop}%
\bibitem [{\citenamefont {Aartsen}\ \emph
  {et~al.}(2014{\natexlab{c}})\citenamefont {Aartsen} \emph
  {et~al.}}]{Aartsen:2014njl}%
  \BibitemOpen
  \bibfield  {author} {\bibinfo {author} {\bibfnamefont {M.~G.}\ \bibnamefont
  {Aartsen}} \emph {et~al.} (\bibinfo {collaboration} {IceCube}),\ }\bibfield
  {title} {\enquote {\bibinfo {title} {{IceCube-Gen2: A Vision for the Future
  of Neutrino Astronomy in Antarctica}},}\ }\href@noop {} {\  (\bibinfo {year}
  {2014}{\natexlab{c}})},\ \Eprint {http://arxiv.org/abs/1412.5106}
  {arXiv:1412.5106 [astro-ph.HE]} \BibitemShut {NoStop}%
\bibitem [{\citenamefont {Kowalski}()}]{Kowalski2018}%
  \BibitemOpen
  \bibfield  {author} {\bibinfo {author} {\bibfnamefont {M}~\bibnamefont
  {Kowalski}},\ }\href@noop {} {\enquote {\bibinfo {title} {{Next-Generation
  IceCube}},}\ }\bibinfo {note} {Presented at Neutrino Oscillation Workshop
  (NOW) 2018, September 12, 2018, Ostuni, Italy.}\BibitemShut {Stop}%
\bibitem [{\citenamefont {Heitler}(1954)}]{Heitler1954}%
  \BibitemOpen
  \bibfield  {author} {\bibinfo {author} {\bibfnamefont {W.}~\bibnamefont
  {Heitler}},\ }\href@noop {} {\emph {\bibinfo {title} {{The Quantum Theory of
  Radiation}}}},\ \bibinfo {edition} {3rd}\ ed.\ (\bibinfo  {publisher} {Dover
  Publications},\ \bibinfo {year} {1954})\BibitemShut {NoStop}%
\bibitem [{\citenamefont {Matthews}(2005)}]{Matthews2005}%
  \BibitemOpen
  \bibfield  {author} {\bibinfo {author} {\bibfnamefont {J.}~\bibnamefont
  {Matthews}},\ }\bibfield  {title} {\enquote {\bibinfo {title} {{A Heitler
  model of extensive air showers}},}\ }\href {\doibase
  10.1016/j.astropartphys.2004.09.003} {\bibfield  {journal} {\bibinfo
  {journal} {Astropart. Phys.}\ }\textbf {\bibinfo {volume} {22}},\ \bibinfo
  {pages} {387--397} (\bibinfo {year} {2005})}\BibitemShut {NoStop}%
\bibitem [{\citenamefont {Lipari}(2009{\natexlab{a}})}]{Lipari:2008td}%
  \BibitemOpen
  \bibfield  {author} {\bibinfo {author} {\bibfnamefont {P.}~\bibnamefont
  {Lipari}},\ }\bibfield  {title} {\enquote {\bibinfo {title} {{The Concepts of
  'Age' and 'Universality' in Cosmic Ray Showers}},}\ }\href {\doibase
  10.1103/PhysRevD.79.063001} {\bibfield  {journal} {\bibinfo  {journal} {Phys.
  Rev. D}\ }\textbf {\bibinfo {volume} {79}},\ \bibinfo {pages} {063001}
  (\bibinfo {year} {2009}{\natexlab{a}})},\ \Eprint
  {http://arxiv.org/abs/0809.0190} {arXiv:0809.0190 [astro-ph]} \BibitemShut
  {NoStop}%
\bibitem [{\citenamefont {Lipari}(2009{\natexlab{b}})}]{Lipari:2009zz}%
  \BibitemOpen
  \bibfield  {author} {\bibinfo {author} {\bibfnamefont {P.}~\bibnamefont
  {Lipari}},\ }\bibfield  {title} {\enquote {\bibinfo {title} {{Universality of
  cosmic ray shower development}},}\ }\href {\doibase
  10.1016/j.nuclphysbps.2009.09.060} {\bibfield  {journal} {\bibinfo  {journal}
  {Nucl. Phys. Proc. Suppl.}\ }\textbf {\bibinfo {volume} {196}},\ \bibinfo
  {pages} {309--318} (\bibinfo {year} {2009}{\natexlab{b}})}\BibitemShut
  {NoStop}%
\bibitem [{\citenamefont {Rott}\ \emph {et~al.}(2013)\citenamefont {Rott},
  \citenamefont {Siegal-Gaskins},\ and\ \citenamefont {Beacom}}]{Rott:2012qb}%
  \BibitemOpen
  \bibfield  {author} {\bibinfo {author} {\bibfnamefont {C.}~\bibnamefont
  {Rott}}, \bibinfo {author} {\bibfnamefont {J.~M.}\ \bibnamefont
  {Siegal-Gaskins}}, \ and\ \bibinfo {author} {\bibfnamefont {J.~F.}\
  \bibnamefont {Beacom}},\ }\bibfield  {title} {\enquote {\bibinfo {title}
  {{New Sensitivity to Solar WIMP Annihilation using Low-Energy Neutrinos}},}\
  }\href {\doibase 10.1103/PhysRevD.88.055005} {\bibfield  {journal} {\bibinfo
  {journal} {Phys.\ Rev.\ D}\ }\textbf {\bibinfo {volume} {88}},\ \bibinfo
  {pages} {055005} (\bibinfo {year} {2013})},\ \Eprint
  {http://arxiv.org/abs/1208.0827} {arXiv:1208.0827 [astro-ph.HE]} \BibitemShut
  {NoStop}%
\bibitem [{\citenamefont {Li}\ and\ \citenamefont {Beacom}(2014)}]{Li:2014sea}%
  \BibitemOpen
  \bibfield  {author} {\bibinfo {author} {\bibfnamefont {S.~W.}\ \bibnamefont
  {Li}}\ and\ \bibinfo {author} {\bibfnamefont {J.~F.}\ \bibnamefont
  {Beacom}},\ }\bibfield  {title} {\enquote {\bibinfo {title} {{First
  calculation of cosmic-ray muon spallation backgrounds for MeV astrophysical
  neutrino signals in Super-Kamiokande}},}\ }\href {\doibase
  10.1103/PhysRevC.89.045801} {\bibfield  {journal} {\bibinfo  {journal}
  {Phys.\ Rev.\ C}\ }\textbf {\bibinfo {volume} {89}},\ \bibinfo {pages}
  {045801} (\bibinfo {year} {2014})},\ \Eprint {http://arxiv.org/abs/1402.4687}
  {arXiv:1402.4687 [hep-ph]} \BibitemShut {NoStop}%
\bibitem [{\citenamefont {Li}\ and\ \citenamefont
  {Beacom}(2015{\natexlab{a}})}]{Li:2015kpa}%
  \BibitemOpen
  \bibfield  {author} {\bibinfo {author} {\bibfnamefont {S.~W.}\ \bibnamefont
  {Li}}\ and\ \bibinfo {author} {\bibfnamefont {J.~F.}\ \bibnamefont
  {Beacom}},\ }\bibfield  {title} {\enquote {\bibinfo {title} {{Spallation
  Backgrounds in Super-Kamiokande Are Made in Muon-Induced Showers}},}\ }\href
  {\doibase 10.1103/PhysRevD.91.105005} {\bibfield  {journal} {\bibinfo
  {journal} {Phys.~Rev.~D}\ }\textbf {\bibinfo {volume} {91}},\ \bibinfo
  {pages} {105005} (\bibinfo {year} {2015}{\natexlab{a}})},\ \Eprint
  {http://arxiv.org/abs/1503.04823} {arXiv:1503.04823 [hep-ph]} \BibitemShut
  {NoStop}%
\bibitem [{\citenamefont {Li}\ and\ \citenamefont
  {Beacom}(2015{\natexlab{b}})}]{Li:2015lxa}%
  \BibitemOpen
  \bibfield  {author} {\bibinfo {author} {\bibfnamefont {S.~W.}\ \bibnamefont
  {Li}}\ and\ \bibinfo {author} {\bibfnamefont {J.~F.}\ \bibnamefont
  {Beacom}},\ }\bibfield  {title} {\enquote {\bibinfo {title} {{Tagging
  Spallation Backgrounds with Showers in Water-Cherenkov Detectors}},}\ }\href
  {\doibase 10.1103/PhysRevD.92.105033} {\bibfield  {journal} {\bibinfo
  {journal} {Phys.\ Rev.\ D}\ }\textbf {\bibinfo {volume} {92}},\ \bibinfo
  {pages} {105033} (\bibinfo {year} {2015}{\natexlab{b}})},\ \Eprint
  {http://arxiv.org/abs/1508.05389} {arXiv:1508.05389 [physics.ins-det]}
  \BibitemShut {NoStop}%
\bibitem [{\citenamefont {Tuli}()}]{Tuli2016}%
  \BibitemOpen
  \bibfield  {author} {\bibinfo {author} {\bibfnamefont {J.~K.}\ \bibnamefont
  {Tuli}},\ }\href@noop {} {\enquote {\bibinfo {title} {{Evaluated Nuclear
  Structure Data File (ENSDF)}},}\ }\bibinfo {howpublished}
  {\url{http://www.nndc.bnl.gov/ensdf/}},\ \bibinfo {note} {[Online; accessed
  2016-05-20]}\BibitemShut {NoStop}%
\bibitem [{\citenamefont {Ferrari}\ \emph {et~al.}(2005)\citenamefont
  {Ferrari}, \citenamefont {Sala}, \citenamefont {Fasso},\ and\ \citenamefont
  {Ranft}}]{Ferrari2005}%
  \BibitemOpen
  \bibfield  {author} {\bibinfo {author} {\bibfnamefont {A.}~\bibnamefont
  {Ferrari}}, \bibinfo {author} {\bibfnamefont {P.~R.}\ \bibnamefont {Sala}},
  \bibinfo {author} {\bibfnamefont {A.}~\bibnamefont {Fasso}}, \ and\ \bibinfo
  {author} {\bibfnamefont {J.}~\bibnamefont {Ranft}},\ }\bibfield  {title}
  {\enquote {\bibinfo {title} {{FLUKA: A multi-particle transport code (Program
  version 2005)}},}\ }\href@noop {} {\  (\bibinfo {year} {2005})}\BibitemShut
  {NoStop}%
\bibitem [{\citenamefont {Battistoni}\ \emph {et~al.}(2007)\citenamefont
  {Battistoni} \emph {et~al.}}]{Battistoni2007}%
  \BibitemOpen
  \bibfield  {author} {\bibinfo {author} {\bibfnamefont {G.}~\bibnamefont
  {Battistoni}} \emph {et~al.},\ }\bibfield  {title} {\enquote {\bibinfo
  {title} {{The FLUKA code: Description and benchmarking}},}\ }\href {\doibase
  10.1063/1.2720455} {\bibfield  {journal} {\bibinfo  {journal} {AIP Conf.
  Proc.}\ }\textbf {\bibinfo {volume} {896}},\ \bibinfo {pages} {31--49}
  (\bibinfo {year} {2007})}\BibitemShut {NoStop}%
\bibitem [{\citenamefont {Abe}\ \emph {et~al.}(2010)\citenamefont {Abe} \emph
  {et~al.}}]{Abe:2009aa}%
  \BibitemOpen
  \bibfield  {author} {\bibinfo {author} {\bibfnamefont {S.}~\bibnamefont
  {Abe}} \emph {et~al.} (\bibinfo {collaboration} {KamLAND}),\ }\bibfield
  {title} {\enquote {\bibinfo {title} {{Production of Radioactive Isotopes
  through Cosmic Muon Spallation in KamLAND}},}\ }\href {\doibase
  10.1103/PhysRevC.81.025807} {\bibfield  {journal} {\bibinfo  {journal} {Phys.
  Rev. C}\ }\textbf {\bibinfo {volume} {81}},\ \bibinfo {pages} {025807}
  (\bibinfo {year} {2010})},\ \Eprint {http://arxiv.org/abs/0907.0066}
  {arXiv:0907.0066 [hep-ex]} \BibitemShut {NoStop}%
\bibitem [{\citenamefont {Bellini}\ \emph {et~al.}(2013)\citenamefont {Bellini}
  \emph {et~al.}}]{Bellini:2013pxa}%
  \BibitemOpen
  \bibfield  {author} {\bibinfo {author} {\bibfnamefont {G.}~\bibnamefont
  {Bellini}} \emph {et~al.} (\bibinfo {collaboration} {Borexino}),\ }\bibfield
  {title} {\enquote {\bibinfo {title} {{Cosmogenic Backgrounds in Borexino at
  3800 m water-equivalent depth}},}\ }\href {\doibase
  10.1088/1475-7516/2013/08/049} {\bibfield  {journal} {\bibinfo  {journal}
  {JCAP}\ }\textbf {\bibinfo {volume} {1308}},\ \bibinfo {pages} {049}
  (\bibinfo {year} {2013})},\ \Eprint {http://arxiv.org/abs/1304.7381}
  {arXiv:1304.7381 [physics.ins-det]} \BibitemShut {NoStop}%
\bibitem [{\citenamefont {Zhang}\ \emph {et~al.}(2016)\citenamefont {Zhang}
  \emph {et~al.}}]{Super-Kamiokande:2015xra}%
  \BibitemOpen
  \bibfield  {author} {\bibinfo {author} {\bibfnamefont {Y.}~\bibnamefont
  {Zhang}} \emph {et~al.} (\bibinfo {collaboration} {Super-Kamiokande}),\
  }\bibfield  {title} {\enquote {\bibinfo {title} {{First measurement of
  radioactive isotope production through cosmic-ray muon spallation in
  Super-Kamiokande IV}},}\ }\href {\doibase 10.1103/PhysRevD.93.012004}
  {\bibfield  {journal} {\bibinfo  {journal} {Phys. Rev. D}\ }\textbf {\bibinfo
  {volume} {93}},\ \bibinfo {pages} {012004} (\bibinfo {year} {2016})},\
  \Eprint {http://arxiv.org/abs/1509.08168} {arXiv:1509.08168 [hep-ex]}
  \BibitemShut {NoStop}%
\bibitem [{\citenamefont {Beacom}\ and\ \citenamefont
  {Vagins}(2004)}]{Beacom:2003nk}%
  \BibitemOpen
  \bibfield  {author} {\bibinfo {author} {\bibfnamefont {J.~F.}\ \bibnamefont
  {Beacom}}\ and\ \bibinfo {author} {\bibfnamefont {M.~R.}\ \bibnamefont
  {Vagins}},\ }\bibfield  {title} {\enquote {\bibinfo {title} {{GADZOOKS!
  Anti-neutrino spectroscopy with large water Cherenkov detectors}},}\ }\href
  {\doibase 10.1103/PhysRevLett.93.171101} {\bibfield  {journal} {\bibinfo
  {journal} {Phys.\ Rev.\ Lett.}\ }\textbf {\bibinfo {volume} {93}},\ \bibinfo
  {pages} {171101} (\bibinfo {year} {2004})},\ \Eprint
  {http://arxiv.org/abs/hep-ph/0309300} {arXiv:hep-ph/0309300 [hep-ph]}
  \BibitemShut {NoStop}%
\bibitem [{\citenamefont {Askins}\ \emph {et~al.}(2015)\citenamefont {Askins}
  \emph {et~al.}}]{Askins:2015bmb}%
  \BibitemOpen
  \bibfield  {author} {\bibinfo {author} {\bibfnamefont {M.}~\bibnamefont
  {Askins}} \emph {et~al.} (\bibinfo {collaboration} {WATCHMAN}),\ }\bibfield
  {title} {\enquote {\bibinfo {title} {{The Physics and Nuclear
  Nonproliferation Goals of WATCHMAN: A WAter CHerenkov Monitor for
  ANtineutrinos}},}\ }\href@noop {} {\  (\bibinfo {year} {2015})},\ \Eprint
  {http://arxiv.org/abs/1502.01132} {arXiv:1502.01132 [physics.ins-det]}
  \BibitemShut {NoStop}%
\bibitem [{\citenamefont {Anghel}\ \emph {et~al.}(2015)\citenamefont {Anghel}
  \emph {et~al.}}]{Anghel:2015xxt}%
  \BibitemOpen
  \bibfield  {author} {\bibinfo {author} {\bibfnamefont {I.}~\bibnamefont
  {Anghel}} \emph {et~al.} (\bibinfo {collaboration} {ANNIE}),\ }\bibfield
  {title} {\enquote {\bibinfo {title} {{Letter of Intent: The Accelerator
  Neutrino Neutron Interaction Experiment (ANNIE)}},}\ }\href@noop {} {\
  (\bibinfo {year} {2015})},\ \Eprint {http://arxiv.org/abs/1504.01480}
  {arXiv:1504.01480 [physics.ins-det]} \BibitemShut {NoStop}%
\bibitem [{\citenamefont {Fernandez}(2016)}]{Fernandez:2016eux}%
  \BibitemOpen
  \bibfield  {author} {\bibinfo {author} {\bibfnamefont {P.}~\bibnamefont
  {Fernandez}} (\bibinfo {collaboration} {Super-Kamiokande}),\ }\bibfield
  {title} {\enquote {\bibinfo {title} {{Status of GADZOOKS!: Neutron Tagging in
  Super-Kamiokande}},}\ }\bibfield  {booktitle} {\emph {\bibinfo {booktitle}
  {{Proceedings, 37th International Conference on High Energy Physics (ICHEP
  2014)}}},\ }\href {\doibase 10.1016/j.nuclphysbps.2015.09.050} {\bibfield
  {journal} {\bibinfo  {journal} {Nucl. Part. Phys. Proc.}\ }\textbf {\bibinfo
  {volume} {273-275}},\ \bibinfo {pages} {353--360} (\bibinfo {year}
  {2016})}\BibitemShut {NoStop}%
\bibitem [{\citenamefont {Aartsen}\ \emph
  {et~al.}(2014{\natexlab{d}})\citenamefont {Aartsen} \emph
  {et~al.}}]{Aartsen:2013vja}%
  \BibitemOpen
  \bibfield  {author} {\bibinfo {author} {\bibfnamefont {M.~G.}\ \bibnamefont
  {Aartsen}} \emph {et~al.} (\bibinfo {collaboration} {IceCube}),\ }\bibfield
  {title} {\enquote {\bibinfo {title} {{Energy Reconstruction Methods in the
  IceCube Neutrino Telescope}},}\ }\href {\doibase
  10.1088/1748-0221/9/03/P03009} {\bibfield  {journal} {\bibinfo  {journal}
  {JINST}\ }\textbf {\bibinfo {volume} {9}},\ \bibinfo {pages} {P03009}
  (\bibinfo {year} {2014}{\natexlab{d}})},\ \Eprint
  {http://arxiv.org/abs/1311.4767} {arXiv:1311.4767 [physics.ins-det]}
  \BibitemShut {NoStop}%
\bibitem [{\citenamefont {Kopper}()}]{Kopper2013}%
  \BibitemOpen
  \bibfield  {author} {\bibinfo {author} {\bibfnamefont {C.}~\bibnamefont
  {Kopper}},\ }\href@noop {} {\enquote {\bibinfo {title} {{Recent IceCube
  Results on High Energy Neutrinos}},}\ }\bibinfo {howpublished}
  {\url{http://macros2013.in2p3.fr/_shared/doc/talks/kopper.pdf}},\ \bibinfo
  {note} {{MACROS 2013}}\BibitemShut {NoStop}%
\bibitem [{\citenamefont {Aartsen}\ \emph
  {et~al.}(2013{\natexlab{c}})\citenamefont {Aartsen} \emph
  {et~al.}}]{Aartsen:2013nla}%
  \BibitemOpen
  \bibfield  {author} {\bibinfo {author} {\bibfnamefont {M.~G.}\ \bibnamefont
  {Aartsen}} \emph {et~al.} (\bibinfo {collaboration} {IceCube}),\ }\bibfield
  {title} {\enquote {\bibinfo {title} {{The IceCube Neutrino Observatory Part
  V: Neutrino Oscillations and Supernova Searches}},}\ }\href@noop {} {\
  (\bibinfo {year} {2013}{\natexlab{c}})},\ \Eprint
  {http://arxiv.org/abs/1309.7008} {arXiv:1309.7008 [astro-ph.HE]} \BibitemShut
  {NoStop}%
\bibitem [{\citenamefont {Aartsen}\ \emph {et~al.}(2017)\citenamefont {Aartsen}
  \emph {et~al.}}]{Aartsen:2016nxy}%
  \BibitemOpen
  \bibfield  {author} {\bibinfo {author} {\bibfnamefont {M.~G.}\ \bibnamefont
  {Aartsen}} \emph {et~al.} (\bibinfo {collaboration} {IceCube}),\ }\bibfield
  {title} {\enquote {\bibinfo {title} {{The IceCube Neutrino Observatory:
  Instrumentation and Online Systems}},}\ }\href {\doibase
  10.1088/1748-0221/12/03/P03012} {\bibfield  {journal} {\bibinfo  {journal}
  {JINST}\ }\textbf {\bibinfo {volume} {12}},\ \bibinfo {pages} {P03012}
  (\bibinfo {year} {2017})},\ \Eprint {http://arxiv.org/abs/1612.05093}
  {arXiv:1612.05093 [astro-ph.IM]} \BibitemShut {NoStop}%
\bibitem [{\citenamefont {Aartsen}\ \emph
  {et~al.}(2015{\natexlab{d}})\citenamefont {Aartsen} \emph
  {et~al.}}]{Aartsen:2015xup}%
  \BibitemOpen
  \bibfield  {author} {\bibinfo {author} {\bibfnamefont {M.~G.}\ \bibnamefont
  {Aartsen}} \emph {et~al.} (\bibinfo {collaboration} {IceCube}),\ }\bibfield
  {title} {\enquote {\bibinfo {title} {{Measurement of the Atmospheric $\nu_e$
  Spectrum with IceCube}},}\ }\href {\doibase 10.1103/PhysRevD.91.122004}
  {\bibfield  {journal} {\bibinfo  {journal} {Phys. Rev. D}\ }\textbf {\bibinfo
  {volume} {91}},\ \bibinfo {pages} {122004} (\bibinfo {year}
  {2015}{\natexlab{d}})},\ \Eprint {http://arxiv.org/abs/1504.03753}
  {arXiv:1504.03753 [astro-ph.HE]} \BibitemShut {NoStop}%
\bibitem [{\citenamefont {K{\"o}epke}\ and\ \citenamefont
  {Steuer}(2016--2017)}]{Koepke:private}%
  \BibitemOpen
  \bibfield  {author} {\bibinfo {author} {\bibfnamefont {L.}~\bibnamefont
  {K{\"o}epke}}\ and\ \bibinfo {author} {\bibfnamefont {A.}~\bibnamefont
  {Steuer}},\ }\href@noop {} {} (\bibinfo {year} {2016--2017}),\ \bibinfo
  {note} {private communications}\BibitemShut {NoStop}%
\bibitem [{\citenamefont {Abbasi}\ \emph {et~al.}(2010)\citenamefont {Abbasi}
  \emph {et~al.}}]{Abbasi:2010vc}%
  \BibitemOpen
  \bibfield  {author} {\bibinfo {author} {\bibfnamefont {R.}~\bibnamefont
  {Abbasi}} \emph {et~al.} (\bibinfo {collaboration} {IceCube}),\ }\bibfield
  {title} {\enquote {\bibinfo {title} {{Calibration and Characterization of the
  IceCube Photomultiplier Tube}},}\ }\href {\doibase
  10.1016/j.nima.2010.03.102} {\bibfield  {journal} {\bibinfo  {journal}
  {Nucl.~Instrum.~Meth.~A}\ }\textbf {\bibinfo {volume} {618}},\ \bibinfo
  {pages} {139--152} (\bibinfo {year} {2010})},\ \Eprint
  {http://arxiv.org/abs/1002.2442} {arXiv:1002.2442 [astro-ph.IM]} \BibitemShut
  {NoStop}%
\bibitem [{\citenamefont {Larson}(2013)}]{Larson2013}%
  \BibitemOpen
  \bibfield  {author} {\bibinfo {author} {\bibfnamefont {M.~J.}\ \bibnamefont
  {Larson}},\ }\emph {\bibinfo {title} {{Simulation and Identification of
  Non-Poissonian Noise Triggers in the IceCube Neutrino Detector}}},\
  \href@noop {} {Master's thesis},\ \bibinfo  {school} {The University of
  Alabama} (\bibinfo {year} {2013})\BibitemShut {NoStop}%
\bibitem [{\citenamefont {Incandela}\ \emph {et~al.}(1988)\citenamefont
  {Incandela} \emph {et~al.}}]{Incandela:1987dh}%
  \BibitemOpen
  \bibfield  {author} {\bibinfo {author} {\bibfnamefont {J.~R.}\ \bibnamefont
  {Incandela}} \emph {et~al.},\ }\bibfield  {title} {\enquote {\bibinfo {title}
  {{The Performance of Photomultipliers Exposed to Helium}},}\ }\href {\doibase
  10.1016/0168-9002(88)90885-6} {\bibfield  {journal} {\bibinfo  {journal}
  {Nucl. Instrum. Meth. A}\ }\textbf {\bibinfo {volume} {269}},\ \bibinfo
  {pages} {237--245} (\bibinfo {year} {1988})}\BibitemShut {NoStop}%
\bibitem [{\citenamefont {Bristow}(2002)}]{Bristow02}%
  \BibitemOpen
  \bibfield  {author} {\bibinfo {author} {\bibfnamefont {M.~P.}\ \bibnamefont
  {Bristow}},\ }\bibfield  {title} {\enquote {\bibinfo {title} {Suppression of
  afterpulsing in photomultipliers by gating the photocathode},}\ }\href
  {\doibase 10.1364/AO.41.004975} {\bibfield  {journal} {\bibinfo  {journal}
  {Appl. Opt.}\ }\textbf {\bibinfo {volume} {41}},\ \bibinfo {pages}
  {4975--4987} (\bibinfo {year} {2002})}\BibitemShut {NoStop}%
\bibitem [{\citenamefont {Katz}(2006)}]{Katz:2006wv}%
  \BibitemOpen
  \bibfield  {author} {\bibinfo {author} {\bibfnamefont {U.~F.}\ \bibnamefont
  {Katz}},\ }\bibfield  {title} {\enquote {\bibinfo {title} {{KM3NeT: Towards a
  km$^3$ Mediterranean Neutrino Telescope}},}\ }\href {\doibase
  10.1016/j.nima.2006.05.235} {\bibfield  {journal} {\bibinfo  {journal} {Nucl.
  Instrum. Meth. A}\ }\textbf {\bibinfo {volume} {567}},\ \bibinfo {pages}
  {457--461} (\bibinfo {year} {2006})},\ \Eprint
  {http://arxiv.org/abs/astro-ph/0606068} {arXiv:astro-ph/0606068 [astro-ph]}
  \BibitemShut {NoStop}%
\bibitem [{\citenamefont {Adrian-Martinez}\ \emph {et~al.}(2016)\citenamefont
  {Adrian-Martinez} \emph {et~al.}}]{Adrian-Martinez:2016fdl}%
  \BibitemOpen
  \bibfield  {author} {\bibinfo {author} {\bibfnamefont {S.}~\bibnamefont
  {Adrian-Martinez}} \emph {et~al.} (\bibinfo {collaboration} {KM3Net}),\
  }\bibfield  {title} {\enquote {\bibinfo {title} {{Letter of intent for KM3NeT
  2.0}},}\ }\href {\doibase 10.1088/0954-3899/43/8/084001} {\bibfield
  {journal} {\bibinfo  {journal} {J.\ Phys.\ G}\ }\textbf {\bibinfo {volume}
  {43}},\ \bibinfo {pages} {084001} (\bibinfo {year} {2016})},\ \Eprint
  {http://arxiv.org/abs/1601.07459} {arXiv:1601.07459 [astro-ph.IM]}
  \BibitemShut {NoStop}%
\bibitem [{\citenamefont {Avrorin}\ \emph {et~al.}(2013)\citenamefont {Avrorin}
  \emph {et~al.}}]{Avrorin:2013sla}%
  \BibitemOpen
  \bibfield  {author} {\bibinfo {author} {\bibfnamefont {A.~V.}\ \bibnamefont
  {Avrorin}} \emph {et~al.},\ }\bibfield  {title} {\enquote {\bibinfo {title}
  {{Current status of the BAIKAL-GVD project}},}\ }\href {\doibase
  10.1016/j.nima.2012.11.151} {\bibfield  {journal} {\bibinfo  {journal} {Nucl.
  Instrum. Meth. A}\ }\textbf {\bibinfo {volume} {725}},\ \bibinfo {pages}
  {23--26} (\bibinfo {year} {2013})}\BibitemShut {NoStop}%
\bibitem [{\citenamefont {Resconi}()}]{Resconi2013}%
  \BibitemOpen
  \bibfield  {author} {\bibinfo {author} {\bibfnamefont {E.}~\bibnamefont
  {Resconi}},\ }\href@noop {} {\enquote {\bibinfo {title} {{The stepping stones
  to proton decay: IceCube, PINGU, MICA}},}\ }\bibinfo {howpublished}
  {\url{https://indico.cern.ch/event/224351/contributions/466197/attachments/368604/513012/Aspen_2013_Resconi.pdf}},\
  \bibinfo {note} {{Aspen Center for Physics, New Directions in Neutrino
  Physics 2013}}\BibitemShut {NoStop}%
\bibitem [{\citenamefont {Steuer}\ and\ \citenamefont
  {K{\"o}pke}(2018)}]{Steuer:2017tca}%
  \BibitemOpen
  \bibfield  {author} {\bibinfo {author} {\bibfnamefont {A.}~\bibnamefont
  {Steuer}}\ and\ \bibinfo {author} {\bibfnamefont {L.}~\bibnamefont
  {K{\"o}pke}},\ }\bibfield  {title} {\enquote {\bibinfo {title} {{Delayed
  light emission to distinguish astrophysical neutrino flavors in IceCube}},}\
  }\bibfield  {booktitle} {\emph {\bibinfo {booktitle} {{Contributions to the
  35th International Cosmic Ray Conference (ICRC 2017)}}},\ }\href {\doibase
  10.22323/1.301.1008} {\bibfield  {journal} {\bibinfo  {journal} {PoS}\
  }\textbf {\bibinfo {volume} {ICRC 2017}},\ \bibinfo {pages} {1008} (\bibinfo
  {year} {2018})}\BibitemShut {NoStop}%
\end{thebibliography}

%


\begin{appendix}

\newpage
\clearpage

\onecolumngrid
\begin{center}
 \bf \large Supplemental Material
\end{center}
\vspace*{0.2cm}


In the main text, we showed how detecting muon echoes can improve discrimination between $\nu_e$-initiated and $\nu_\tau$-initiated showers. We showed results using 100-TeV showers and a flavor composition of the form $( x : 1-2x : x)_\oplus$.  Here we provide more details on the statistical method and show how the results depend on choices of inputs.

In Appendix~\ref{sec:formalism}, we present the underlying formalism for flavor discrimination per shower.  In Appendix~\ref{sec:sensitivity}, we apply it to an ensemble of showers.  In Appendix~\ref{sec:muon_other_energies}, we discuss flavor discrimination at other shower energies.  In Appendix~\ref{sec:neutron}, we discuss neutron echoes.  In Appendix~\ref{sec:other_assumptions}, we show sensitivity results for other input choices.

To simplify the notation, we explicitly show the shower energy $E_\text{sh}$ dependence when defining a quantity, and suppress it otherwise.  In the probability definitions, we show $\nu$ CC cases explicitly; NC and $\bar\nu$ cases have similar definitions, with CC replaced by NC, and $\nu$ replaced by $\bar{\nu}$.


\section{Flavor discrimination for one shower}
\label{sec:formalism}

We calculate the probability that an observed shower, containing $N_\mu$ muon decays, was initiated by a neutrino $\nu_l$, of definite flavor $l = e, \mu$, or $\tau$.

The main observable of a shower is its measured energy $E_\text{sh}$, which is proportional to the total collected light.  This is generally different from the true shower energy because of the detector energy resolution.  Because the resolution is narrow, we simply assume that the true shower energy is in the range $\left[ 0.9, 1.1\right] E_\text{sh}$.  This mimics the effect of having a resolution of about 10\% in $E_\text{sh}$.

Using Bayes' theorem, the probability that a shower with energy $E_\text{sh}$ and $N_\mu$ muon decays was initiated by a $\nu_l$ is
%
 \begin{equation}
  \label{equ:ProbFlav1}
  P_{\nu_l|N_\mu} \left(E_\text{sh}\right)
  = \frac{ P_{N_\mu|\nu_l}^\text{CC} P_{\nu_l}^\text{CC} + P_{N_\mu|\nu_l}^\text{NC} P_{\nu_l}^\text{NC} }
         { \sum_{\alpha = e,\mu,\tau}
         \left[ \left( P_{N_\mu|\nu_\alpha}^\text{CC} P_{\nu_\alpha}^\text{CC} + P_{N_\mu|\bar{\nu}_\alpha}^\text{CC} P_{\bar{\nu}_\alpha}^\text{CC} \right) +
                                                     \left( P_{N_\mu|\nu_\alpha}^\text{NC} P_{\nu_\alpha}^\text{NC} + P_{N_\mu|\bar{\nu}_\alpha}^\text{NC} P_{\bar{\nu}_\alpha}^\text{NC} \right) \right] } \;.
 \end{equation}
%
Here, $P_{\nu_l}^{\text{CC}}\left(E_\text{sh}\right)$ is the probability that a shower with energy $E_\text{sh}$ is produced by the CC interaction of a $\nu_l$, which we detail below, while $P_{N_\mu|\nu_l}^{\text{CC}}\left(E_\text{sh}\right)$ is the probability that said shower yields $N_\mu$ muon decays, which is calculated via \texttt{FLUKA} simulations and shown in Figs.~4 and \ref{fig:muon_decay_energies} for different shower energies.

The probability $P_{\nu_l}^{\text{CC}}$ is defined as
\begin{equation}
 \label{equ:ProbNuADef} 
 P_{\nu_l}^{\text{CC}}\left(E_\text{sh}\right)
 =
 \frac{ N_{\nu_l}^{\text{CC}} }
      { \sum_{\alpha = e,\mu,\tau}  \left( N_{\nu_\alpha}^\text{CC} + N_{\bar{\nu}_\alpha}^\text{CC} \right)
                           + \left( N_{\nu_\alpha}^\text{NC}  + N_{\bar{\nu}_\alpha}^\text{NC} \right) } \;,
\end{equation}
where $N_{\nu_l}^{\text{CC}}\left(E_\text{sh}\right)$ is the number of $\nu_l$-initiated showers generated by CC interactions.  The denominator in Eq.~(\ref{equ:ProbNuADef}) is the total number of showers initiated by all flavors of neutrinos and anti-neutrinos.

To calculate the number of showers, we use the ``theorist's approach''~\cite{Laha:2013lka}, assuming perfect detector efficiency at the relevant energies.  The final results on flavor discrimination are affected by only the relative, not the absolute, event rates from different flavors.  We consider a flux $F_{\nu_l}$ of $\nu_l$ (in units of GeV$^{-1}$ cm$^{-2}$ s$^{-1}$ sr$^{-1}$) arriving at the detector, which contains $\mathcal{N}$ target nucleons.  The flux already includes any attenuation due to propagation in the Earth.  In observation time $\Delta t$, within a neutrino solid angle $\Delta \Omega$, the number of detected $\nu_l$-initiated CC showers in an energy bin is
%
 \begin{equation}
  \label{equ:NumberOfShowers1}
  N_{\nu_l}^\text{CC}\left(E_\text{sh}\right)
  = \mathcal{N} \cdot \Delta t \cdot \Delta \Omega \cdot \int_0^\infty F_{\nu_l}\left(E_\nu\right) \cdot \sigma_{\nu}^\text{CC}\left(E_\nu\right) \cdot g_{\nu_l}^\text{CC}\left(E_\nu,E_\text{sh}\right) dE_\nu \;,
 \end{equation}
%
where $E_\nu$ is the neutrino energy and $\sigma_{\nu}^\text{CC}$ is the neutrino-nucleon CC cross section~\cite{Gandhi:1995tf, Gandhi:1998ri, Connolly:2011vc}.  The function $g_{\nu_l}^\text{CC}$ is the probability that a neutrino with energy $E_\nu$ creates a shower with energy $E_\text{sh}$; it is different for each flavor.

\begin{itemize}[leftmargin=*]

\item
In $\nu_e$ CC interactions, all of the neutrino energy is deposited in the electromagnetic and hadronic showers. Accordingly, we define 
\begin{equation}
 g_{\nu_e}^\text{CC} = \left \{
 \begin{aligned}
  &1, && \text{if}\ E_\nu \in \left[ 0.9, 1.1\right] E_\text{sh} \\
  &0, && \text{otherwise}
 \end{aligned} \right. \;.
\end{equation}

\item
In $\nu_\tau$ CC interactions, the outgoing tau has numerous decay modes.  All of them have outgoing neutrinos, which carry away energy and do not appear in the shower, so that $E_\text{sh} \lesssim E_\nu$.  On average, the outgoing neutrinos carry away 40\% of the tau energy, or 25\% of the primary neutrino energy.  For simplicity, we make $g_{\nu_\tau}^\text{CC}$ nonzero only in the energy range $E_\nu \in [0.9, 1.1] E_\text{sh} / 0.75$.  Since 17\% of tau decays are into muons and neutrinos, without a shower, we estimate
\begin{equation}
 g_{\nu_\tau}^\text{CC} = \left \{
 \begin{aligned}
  &0.83, && \text{if}\ E_\nu \in \left[ 0.9, 1.1\right] E_\text{sh}/0.75 \\
  &0, && \text{otherwise}
 \end{aligned} \right. \;.
\end{equation}

\item
In NC interactions, the energy deposited in the shower is the energy of the final-state hadrons, \ie, $E_\text{sh} = yE_\nu$.  For the shower energy to lie within 10\% of $E_\text{sh}$, the value of $y$ must lie in the range $[y_{\min}, y_{\max}]$, where $y_{\min} \equiv 0.9E_\text{sh}/E_\nu$
 and $y_{\max} \equiv \min\{1.1E_\text{sh}/E_\nu, 1.0\}$.  Hence, we define
\begin{equation}
 g_{\nu_l}^\text{NC}(E_\nu) = \frac{\int^{y_{\max}}_{y_{\min}} \frac{d\sigma_\nu^\text{NC}}{dy}(E_\nu,y)~dy}{\int^1_0 \frac{d\sigma_\nu^\text{NC}}{dy}(E_\nu,y)~dy} \;,
\end{equation}
where $d\sigma_\nu^\text{NC}/dy$ is the $y$ probability distribution for NC interactions~\cite{Connolly:2011vc}.  However, because hadron-initiated showers carry a small fraction $y$ of the neutrino energy, and because the neutrino flux is steeply falling, NC showers are subdominant to CC showers~\cite{Beacom:2004jb}.
\item
In $\nu_\mu$ CC interactions, the outgoing muon leaves an identifiable track.  We exclude these events by setting
\begin{equation}
    g_{\nu_\mu}^\text{CC} = 0 \;.
\end{equation}
We have assumed that no track is misidentified as a shower; otherwise, the value of $g_{\nu_\mu}^\text{CC}$ would be set to the probability of mis-identification.  As with NC events, these would be subdominant in the shower spectrum.
\end{itemize}

We write Eqs.~(\ref{equ:ProbFlav1})--(\ref{equ:NumberOfShowers1}) in a more useful way.  Consider an all-flavor astrophysical neutrino flux $\propto E_\nu^{-\gamma}$ and flavor ratios at Earth $\left( f_{e,\oplus}:f_{\mu,\oplus}:f_{\tau,\oplus} \right)$, such that the flux of $\nu_l$ is $F_{\nu_l} = f_{l,\oplus} F_0 E_\nu^{-\gamma}$, with $F_0$ the normalization of the flux.  With this, \equ{NumberOfShowers1} becomes
\begin{equation}
 \label{equ:NumberOfShowers2}
 N_{\nu_l}^\text{CC}\left(E_\text{sh}\right)
 = \mathcal{N} \cdot \Delta t \cdot \Delta \Omega \cdot F_0 \cdot f_{l,\oplus} \cdot I_{\nu_l}^{\text{CC}}\left(E_\text{sh}\right) \;,
\end{equation}
with the shorthand
 \begin{equation}
  \label{equ:IntegralDef}
  I_{\nu_l}^{\text{CC}}\left(E_\text{sh}\right) \equiv 
  \int_0^\infty E_\nu^{-\gamma} \cdot \sigma_{\nu}^\text{CC}\left(E_\nu\right) \cdot g_{\nu_l}^\text{CC}\left(E_\nu,E_\text{sh}\right) dE_\nu \;.
 \end{equation}
Finally, using Eqs.~(\ref{equ:NumberOfShowers2}) and (\ref{equ:IntegralDef}), and assuming equal flavor ratios for neutrinos and anti-neutrinos, \equ{ProbFlav1} becomes
 \begin{equation}
  \label{equ:ProbFlav2}
  P_{\nu_l|N_\mu} \left(E_\text{sh}\right)
  = \frac{ f_{l,\oplus} \left[ P_{N_\mu|\nu_l}^\text{CC} I_{\nu_l}^\text{CC} + P_{N_\mu|\nu_l}^\text{NC} I_{\nu_l}^\text{NC} \right] }
         { \sum_{\alpha = e,\mu,\tau}
         f_{\alpha,\oplus} \left[ \left( P_{N_\mu|\nu_\alpha}^\text{CC} I_{\nu_\alpha}^\text{CC} + P_{N_\mu|\bar{\nu}_\alpha}^\text{CC} I_{\bar{\nu}_\alpha}^\text{CC} \right) +
                                                                      \left( P_{N_\mu|\nu_\alpha}^\text{NC} I_{\nu_\alpha}^\text{NC} + P_{N_\mu|\bar{\nu}_\alpha}^\text{NC} I_{\bar{\nu}_\alpha}^\text{NC} \right) \right] } \;.
 \end{equation}
The probability that the shower with $N_\mu$ muon decays was created by a $\nu_l$ or a $\bar{\nu}_l$ is simply $P_{\nu_l|N_\mu} + P_{\bar{\nu}_l|N_\mu}$.

\setcounter{figure}{0}
\renewcommand{\thefigure}{A\arabic{figure}}
 \begin{figure}[t!]
  \begin{minipage}{0.48\textwidth}  
    \centering
    \includegraphics[width=\textwidth]{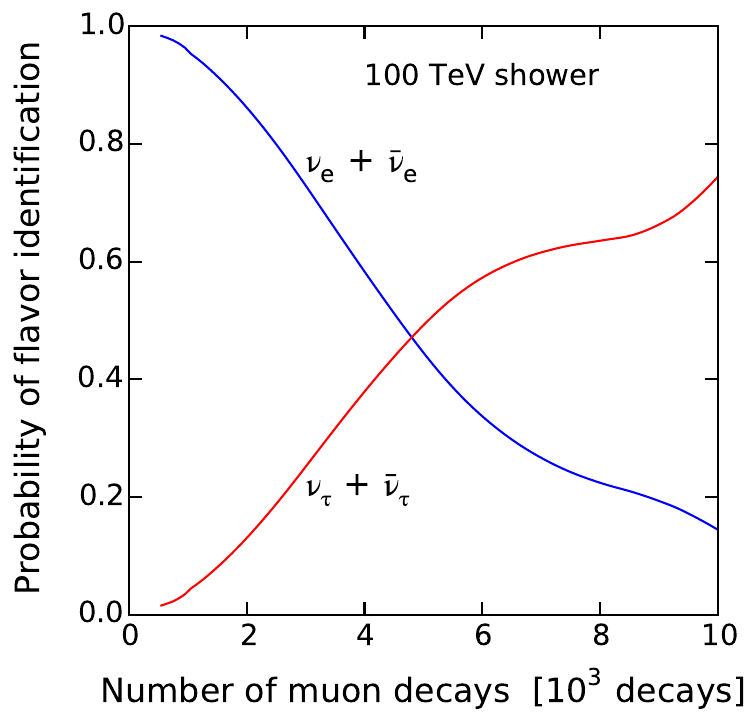}
    \caption{Probability that one neutrino-induced shower was generated by a $\nu_e$ (via either CC or NC) or $\nu_\tau$, as a function of number of muon decays.  The curve for $\nu_\mu$ is calculated but not shown.}
    \label{fig:nu_flavor}
  \end{minipage}
  \hspace*{0.2cm}
  \begin{minipage}{0.48\textwidth}  
   \centering
   \includegraphics[width=\textwidth]{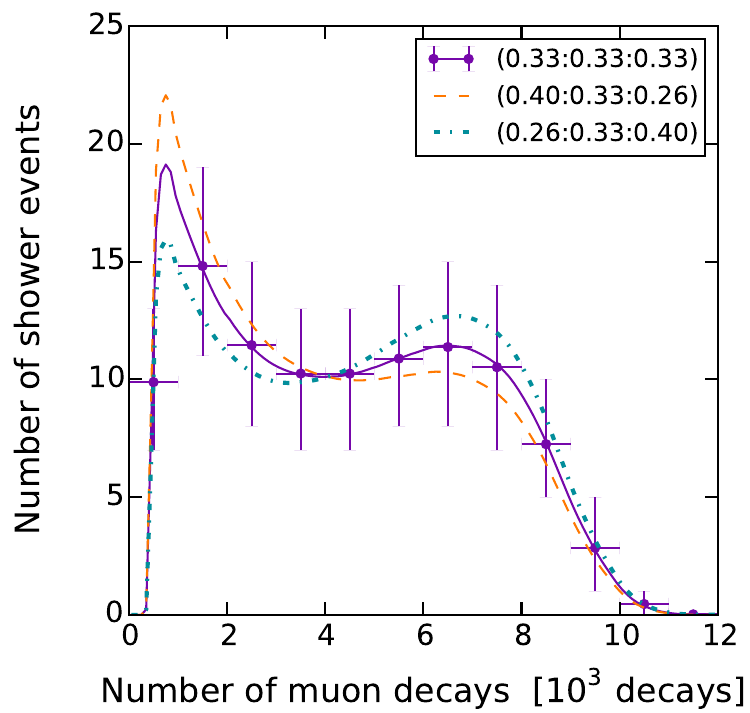}
   \caption{Distributions of muon decays for an ensemble of 100 showers of 100 TeV, for different choices of flavor composition $( f_{e,\oplus} : f_{\mu,\oplus} : f_{\tau,\oplus} )$, reflecting the central value and width of the band in Fig.~1.}
   \label{fig:parameter_fit}
  \end{minipage}
 \end{figure}
%

Figure~\ref{fig:nu_flavor} shows this probability computed at $E_\text{sh} = 100$~TeV, assuming a diffuse astrophysical neutrino flux with spectral index $\gamma = 2.5$ and a flavor composition of $\left( \frac{1}{3} : \frac{1}{3} : \frac{1}{3} \right)_\oplus$, compatible with IceCube results~\cite{Aartsen:2015ita}.  The neutrino is more likely to be a $\nu_e$ if there are fewer muon decays and a $\nu_\tau$ if there are more decays.  The probability that the shower is from a $\nu_\mu$ NC interaction (not shown) reaches at most 10\%, at large values of $N_\mu$.


\section{Flavor discrimination for an ensemble of showers}
\label{sec:sensitivity}

We use the results from Appendix \ref{sec:formalism} to infer the $f_{e,\oplus}$ and $f_{\tau,\oplus}$ flavor ratios of an ensemble of showers.  We first explain how we generate the artificial shower ensemble; then we show how to infer their flavor ratios.

To generate an ensemble of showers with energy $E_\text{sh}$, we first assume a neutrino flux with spectral index $\gamma = 2.5$ and ``real'' values for the flavor ratios $\left( f_{e,\oplus}^\text{r} : f_{\mu,\oplus}^\text{r} : f_{\tau,\oplus}^\text{r} \right)$.  We then use the probability distribution functions of the number of muon decays for each channel, $P_{N_\mu|\nu_l}^{\text{CC}}$ and $P_{N_\mu|\nu_l}^{\text{NC}}$ (shown in Figs.~4 and~\ref{fig:muon_decay_energies}), to construct the total probability distribution of muon decays associated to that flux, by summing over all flavors and interaction channels:
\begin{eqnarray}
 P_{\mu,\text{tot}} \left( N_\mu; E_\text{sh}, \left\{f_{l,\oplus}^\text{r} \right\} \right)
 =
 \sum_{\alpha=e,\mu,\tau} \left[
                           P_{N_\mu|\nu_\alpha}^{\text{CC}}
                           \cdot P_{\nu_\alpha}^{\text{CC}} \left(f_{\alpha,\oplus}^\text{r}\right) +
                           P_{N_\mu|\nu_\alpha}^{\text{NC}} \cdot P_{\nu_\alpha}^{\text{NC}} \left(f_{\alpha,\oplus}^\text{r}\right)
                          \right] \nonumber  +
                          \left[ \nu_\alpha \to \bar{\nu}_\alpha \right] \;.                   
\end{eqnarray}

Figure~\ref{fig:parameter_fit} shows the total muon decay distribution for $E_\text{sh} = 100$ TeV, for three choices of flavor composition.  The distribution for our nominal case $\left( \frac{1}{3} : \frac{1}{3} : \frac{1}{3} \right)_\oplus$ has a saddle shape, peaked at low number of decays due to the sharp distribution of the $\nu_e$ CC channel.  The height of this peak increases with $f_{e,\oplus}$.

We use the above distribution to randomly sample the number of muon decays for each shower, that is, we obtain $N_{\mu,i}$ for $i = 1, \ldots, N_\text{sh}$.  These are our ``real'' data.  We choose $N_\text{sh} = 100$, consistent with near-future expectations for IceCube.  

Merely to illustrate the flavor separation power given this sample size, we include binned data for this choice in Fig.~\ref{fig:parameter_fit}.  The points and error bars show the expected number of showers per bin and its $1\sigma$ Poissonian fluctuation.  The power of flavor discrimination using muon echoes hinges on the ratio of the number of showers with few muon decays --- \eg, $N_\mu < 3000$ --- to the number of showers with many muon decays --- \eg, $N_\mu > 6000$.  A higher ratio drives $f_{e,\oplus}$ up and $f_{\tau,\oplus}$ down, and vice versa.

For our actual analysis, we recover the flavor ratios $f_{l,\oplus}^\text{r}$ of the ensemble via an unbinned maximum likelihood approach.  The likelihood function for a test flavor composition $\left( f_{e,\oplus} : f_{\mu,\oplus} : f_{\tau,\oplus} \right)$, with $f_{\mu,\oplus}\equiv 1-f_{e,\oplus}-f_{\tau,\oplus}$, is
\begin{equation}
\label{eq:likelihood}
 \mathcal{L} \left( f_{e,\oplus}, f_{\tau,\oplus} \right)
 = G(f_{\mu,\oplus})\prod_{i = 1}^{N_\text{sh}} P_{\mu,\text{tot}} \left( N_{\mu,i} ; f_{e,\oplus}, f_{\tau,\oplus} \right) \;.
\end{equation}
The Gaussian term, $G(f_{\mu,\oplus})$, constrains the muon component from deviating too much from its true value, assuming it can be measured from a separate track analysis.  We choose the $1\sigma$ width of the Gaussian to be 0.12, consistent with the present IceCube measurement~\cite{Aartsen:2015ita}.

The maximum value of the likelihood determines the best-fit values of $f_{e,\oplus}$, $f_{\tau,\oplus}$, and $f_{\mu,\oplus} = 1 - f_{e,\oplus} - f_{\tau,\oplus}$.  To estimate the uncertainty on this value, we repeat the maximum likelihood procedure using 1000 random realizations of the real data.  Figure~1 shows the best-fit values and uncertainties on $f_{e,\oplus}$ and $f_{\tau,\oplus}$ that result from this procedure, assuming ensembles of $N_\text{sh} = 100$ showers each and real flavor ratios $f_{e,\oplus}^\text{r} = f_{\tau,\oplus}^\text{r} = ( 1-f_{\mu,\oplus}^\text{r} )/2$, with $f_{\mu,\oplus}^\text{r}$ varying in the range $\left[ 0, 1 \right]$.


\section{Results for different energies}
\label{sec:muon_other_energies}

\begin{figure*}[t]
    \begin{center}                  
        \includegraphics[width=0.48\columnwidth]{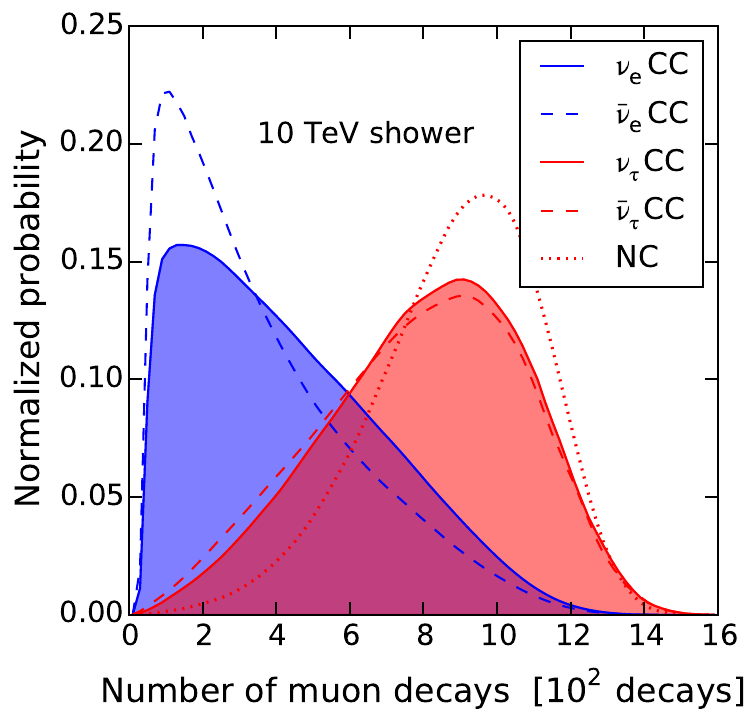}
        \hspace{0.25cm}
        \includegraphics[width=0.48\columnwidth]{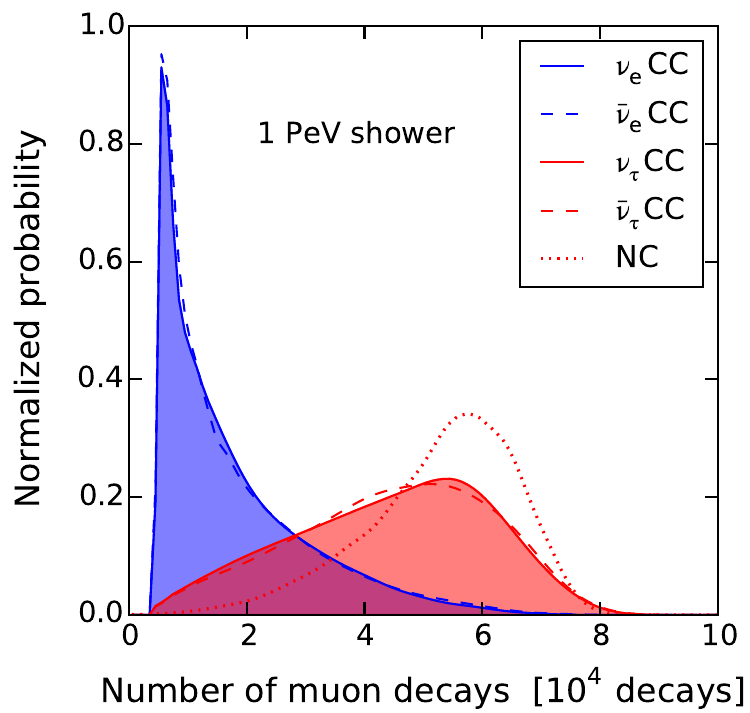}
        \caption{Normalized distributions of the numbers of muon decays per shower of energy 10~TeV and 1~PeV for different neutrino interaction channels.  Note the changes in $x$-axis scale compared to Fig.~4.}
        \label{fig:muon_decay_energies}
    \end{center}
\end{figure*}

In the main text, we consider showers of 100~TeV; the normalized distribution of number of muon decays for this shower energy is shown in Fig.~4.

Figure~\ref{fig:muon_decay_energies} shows the distributions at 10~TeV and 1~PeV.  The same general shapes and behavior of the curves is seen at all energies: $\nu_e$-initiated CC showers have appreciably fewer muon decays than $\nu_\tau$-initiated CC showers and NC showers.  The main change is in the intensity of the muon echo, which scales roughly linearly with shower energy.

As the shower energy changes, there are moderate changes in the results.  The value of $\langle y \rangle$ decreases with increasing energy, which means that $\nu_e$-initiated CC showers become more leptonic.  And the $y$ distributions for $\nu$ and $\bar{\nu}$ become more similar at higher energies, and, therefore, so do their muon decay distributions.  Therefore, the separation between $\nu_e$ and $\nu_\tau$ becomes cleaner at higher energies.  This is evidenced by contrasting the panels in Fig.~\ref{fig:muon_decay_energies}.


\section{Results for neutron echoes}
\label{sec:neutron}

\begin{figure}[t]
    \centering                  
    \includegraphics[width=0.48\columnwidth]{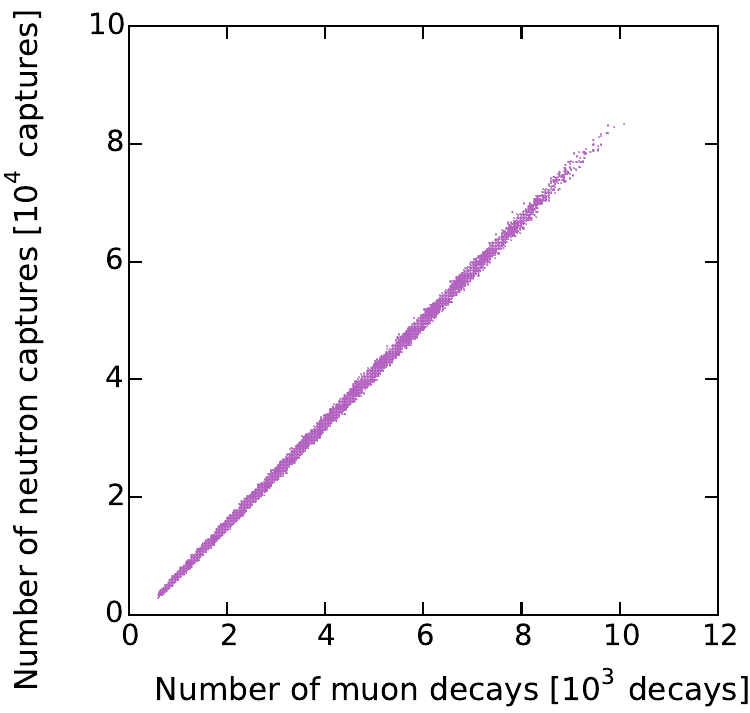}
    \vspace{-6pt}
    \caption{Correlations between numbers of muon decays and neutron captures for individual 100~TeV $\nu_e$ CC shower. Note the different axis scales.}
    \label{fig:neutron_capture}
\end{figure}

Like the muon echo, the neutron echo is a product of the hadronic component of a shower.

Figure~\ref{fig:neutron_capture} shows that the number of muon decays and the number of neutron captures is tightly correlated on an event-by-event basis.  Because of this, the probability distributions of the numbers of neutron captures behave similarly to those of muon decays (Figs.~4 and~\ref{fig:muon_decay_energies}), except for a scaling of the $x$-axis by a factor of about 10.

If we were to incorporate neutron echoes in our sensitivity estimate, Eq.~(\ref{eq:likelihood}) would have an extra term $\prod_{i = 1}^{N_\text{sh}} P_{n,\text{tot}} \left( N_{n,i} ; f_{e,\oplus}, f_{\tau,\oplus} \right)$ on the right-hand side, with $N_{n,i}$ the number of neutron captures in each shower of the ensemble.  However, the distribution of number of neutron captures, $P_{n,\text{tot}}$, is essentially just $P_{\mu,\text{tot}}$ scaled up by a factor of 10.  Therefore, adding it to the likelihood would not alter the best-fit values of $f_{e,\oplus}$ and $f_{\tau,\oplus}$ or their uncertainties.

This is true from a theoretical perspective.  However, from an experimental perspective, neutron echoes are attractive because there seems to be less PMT afterpulsing at late times.

Finally, there is a third possible post-shower signal --- the spallation echo --- coming from the collective Cherenkov light from beta decays of long-lived ($\sim 0.1$--10 s) unstable nuclei.  These isotopes, which are a background in low-energy neutrino detectors, are produced more efficiently in hadronic than electromagnetic showers, by a factor $\sim 10$~\cite{Li:2014sea, Li:2015kpa, Li:2015lxa}.  While the spallation echo is not observable in IceCube or similar detectors due to ambient backgrounds, it might have an application in another context.


\section{Results for other input choices}
\label{sec:other_assumptions}

\begin{figure}[t]
    \centering                
        \includegraphics[width=0.48\textwidth,clip=true,trim = 0 0 0 0.55cm]{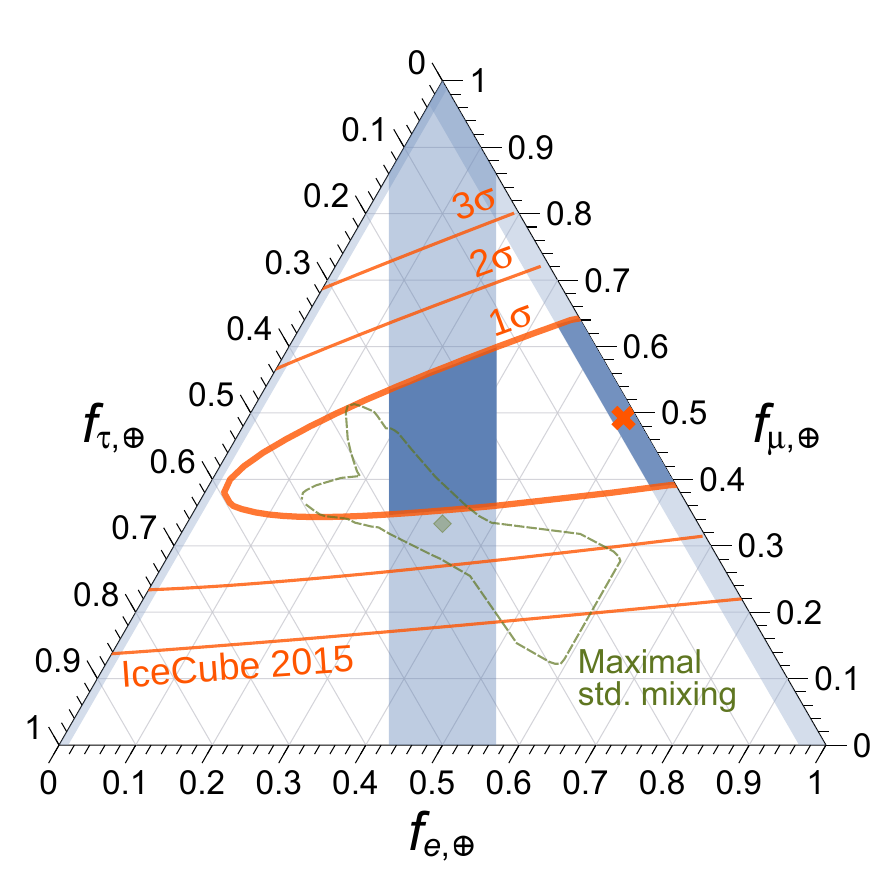}
        \caption{Expected precision of our proposed technique for 100 detected showers at 100~TeV, for different assumptions of the flavor composition $f_{l,\oplus}$ ($l = e,\mu,\tau$) of astrophysical neutrinos at Earth: $(0 : 2x : 1-2x )_\oplus$ (left band), $( x : 1-2x : x )_\oplus$ (central band, same as in Fig.~1), and $( 1-2x : 2x : 0 )_\oplus$ (right band), with $x \in [0,0.5]$.}
        \label{fig:flavor_triangle_flav_assump}
\end{figure}

Figure~\ref{fig:flavor_triangle_flav_assump} shows the flavor sensitivity, using muon echoes, for three different assumptions of the flavor composition at Earth, including the one shown in Fig.~1. 

For the choice of flavor composition in Fig.~1, the average $1\sigma$ uncertainty was 0.07.  For $( 0 : 2x : 1-2x )_\oplus$, with $x \in [0,0.5]$, the best-fit values lie on the left axis of the plot; only the one-sided $1\sigma$ range, of size $0.01$, is visible.  For $( 1-2x : 2x : 0 )_\oplus$, the best-fit values lie on the right axis of the plot; the one-sided $1\sigma$ range, of size $0.04$, is visible.  These are two extreme choices.  Their smaller uncertainties are due to the fact that the total distribution of muon decays of the shower ensemble is dominated by the distribution from either $\nu_e$-initiated or $\nu_\tau$-initiated CC showers.  Hence, our nominal choice of flavor composition, in Fig.~1, was conservative, as it has the largest uncertainty.

At fixed shower energy, the uncertainty on the $\nu_e$ fraction scales as $\sqrt{N_\text{sh}}$, subject to some caveats.  When $N_\text{sh}$ is small ($\lesssim 20$), the likelihood is basically flat, and one typically cannot break the $\nu_e$-$\nu_\tau$ degeneracy with good precision.  When $N_\text{sh}$ is large ($\gtrsim 1000$), one should take a narrower prior on the $\nu_\mu$ fraction to reflect its measurement being correspondingly better.

The flavor sensitivity is robust against other input choices.  For example, the average $1\sigma$ uncertainty is virtually unaffected for a harder neutrino flux of $\gamma = 2$, compared to $\gamma = 2.5$.


\end{appendix}


\end{document}